\documentclass[amsmath,amssymb,aps,prx,reprint,superscriptaddress]{revtex4-2}

\usepackage{amsmath}
\usepackage{xcolor}
\usepackage{amsfonts,amssymb}
\usepackage{graphicx}
\usepackage{hyperref}
\usepackage{xurl}
\usepackage{academicons}
\usepackage{orcidlink}
\usepackage{multirow}
\usepackage{chemformula}
\usepackage{booktabs}

\begin{document}

\title{Harnessing AtomisticSkills for Agentic Atomistic Research}

\author{Bowen Deng\,\orcidlink{0000-0003-4085-381X}}
\affiliation{Department of Materials Science and Engineering, Massachusetts Institute of Technology, Cambridge, MA 02139, USA}
\author{Bohan Li\,\orcidlink{0009-0000-9946-630X}}
\altaffiliation{These authors contributed equally as co-second authors.}
\affiliation{Department of Materials Science and Engineering, Massachusetts Institute of Technology, Cambridge, MA 02139, USA}
\author{Matthew Cox\,\orcidlink{0009-0008-9482-4217}}
\altaffiliation{These authors contributed equally as co-second authors.}
\affiliation{Department of Chemical Engineering, Massachusetts Institute of Technology, Cambridge, MA 02139, USA}
\author{Hoje Chun\,\orcidlink{0000-0003-0624-536X}}
\affiliation{Department of Materials Science and Engineering, Massachusetts Institute of Technology, Cambridge, MA 02139, USA}
\affiliation{Department of Chemistry, Kookmin University, Seoul 02707, Republic of Korea}
\author{Juno Nam\,\orcidlink{0000-0002-9506-2938}}
\affiliation{Department of Materials Science and Engineering, Massachusetts Institute of Technology, Cambridge, MA 02139, USA}
\author{Artur Lyssenko\,\orcidlink{0009-0003-0863-2678}}
\affiliation{Harvard University, Department of Chemistry and Chemical Biology, Cambridge, MA, 02138, USA}
\affiliation{Department of Materials Science and Engineering, Massachusetts Institute of Technology, Cambridge, MA 02139, USA}
\author{Sathya Edamadaka\,\orcidlink{0000-0002-3477-7825}}
\affiliation{Department of Materials Science and Engineering, Massachusetts Institute of Technology, Cambridge, MA 02139, USA}
\author{Jurgis Ruza\,\orcidlink{0009-0008-7151-1250}}
\affiliation{Department of Materials Science and Engineering, Massachusetts Institute of Technology, Cambridge, MA 02139, USA}
\author{Xiaochen Du\,\orcidlink{0000-0001-6228-0907}}
\affiliation{Department of Chemical Engineering, Massachusetts Institute of Technology, Cambridge, MA 02139, USA}
\author{Nofit Segal\,\orcidlink{0000-0002-8891-8590}}
\affiliation{Department of Materials Science and Engineering, Massachusetts Institute of Technology, Cambridge, MA 02139, USA}
\author{Jesus Diaz Sanchez\,\orcidlink{0000-0001-6068-318X}}
\affiliation{Department of Chemistry, Massachusetts Institute of Technology, Cambridge, MA 02139, USA}
\author{Mingrou Xie\,\orcidlink{0000-0002-1564-586X}}
\affiliation{Department of Chemical Engineering, Massachusetts Institute of Technology, Cambridge, MA 02139, USA}
\author{Ty Perez\,\orcidlink{0000-0002-8590-533X}}
\affiliation{Department of Materials Science and Engineering, Massachusetts Institute of Technology, Cambridge, MA 02139, USA}
\author{Yu Yao\,\orcidlink{0009-0000-7382-878X}}
\affiliation{Department of Nuclear Science and Engineering, Massachusetts Institute of Technology, Cambridge, MA 02139, USA}
\author{Miguel Steiner\, \orcidlink{0000-0002-7634-7268}}
\affiliation{Department of Materials Science and Engineering, Massachusetts Institute of Technology, Cambridge, MA 02139, USA}
\author{Sauradeep Majumdar\, \orcidlink{0000-0002-2095-3082}}
\affiliation{Department of Materials Science and Engineering, Massachusetts Institute of Technology, Cambridge, MA 02139, USA}
\author{Charles B. Musgrave III\, \orcidlink{0000-0002-3432-0817}}
\affiliation{Department of Materials Science and Engineering, Massachusetts Institute of Technology, Cambridge, MA 02139, USA}
\author{Anirban Chandra\, \orcidlink{0000-0002-9120-3051}}
\affiliation{Shell Information Technology International Inc., Texas 77082, United States}
\author{Abhirup Patra\, \orcidlink{0000-0002-2446-8017}}
\affiliation{Shell International Exploration \& Production Inc., Texas 77079, United States}
\author{Detlef Hohl\, \orcidlink{0000-0002-9737-5524}}
\affiliation{Shell Information Technology International Inc., Texas 77082, United States}
\author{Connor W. Coley\,\orcidlink{0000-0002-8271-8723}}
\affiliation{Department of Chemical Engineering, Massachusetts Institute of Technology, Cambridge, MA 02139, USA}
\author{Ju Li\,\orcidlink{0000-0002-7841-8058}}
\affiliation{Department of Materials Science and Engineering, Massachusetts Institute of Technology, Cambridge, MA 02139, USA}
\affiliation{Department of Nuclear Science and Engineering, Massachusetts Institute of Technology, Cambridge, MA 02139, USA}
\author{Rafael Gómez-Bombarelli\,\orcidlink{0000-0002-9495-8599}}
\email[]{rafagb@mit.edu}
\affiliation{Department of Materials Science and Engineering, Massachusetts Institute of Technology, Cambridge, MA 02139, USA}

\date{\today}

\begin{abstract}
Computational materials science and chemistry span vast knowledge domains and fractured software ecosystems. Although large language models (LLMs) have demonstrated research capabilities, scaling monolithic agents to manage the rigor and complexity of atomistic research remains a challenge. Here, we introduce AtomisticSkills, an open-source harness framework that empowers general-purpose AI coding agents to conduct atomistic research across materials science, chemistry, and drug discovery. By hierarchically decomposing scientific workflows into agent skills and tools, AtomisticSkills provides agents with modular, extensible, and plug-and-play research capabilities. The framework integrates more than 100 human-curated multidisciplinary skills, including database access, thermodynamics and kinetics modeling, and diverse simulation engines employing machine learning interatomic potentials (MLIPs) and density functional theory (DFT). We validate its functional coverage against scientific literature and demonstrate robust orchestration capabilities across diverse scientific campaigns: generative design of Li-ion solid-state electrolytes, high-throughput screening of metal-organic frameworks for \ch{CO2} capture, autonomous MLIP benchmarking and fine-tuning, multi-stage structure-based virtual screening for drug design, multimodal X-ray diffraction pattern analysis, and screening of Fe-oxide catalysts for oxygen evolution reaction. AtomisticSkills provides a critical agent infrastructure towards building fully autonomous AI scientists.
\end{abstract}


\maketitle

\section{Introduction}

\begin{figure*}[th]
\centering
\includegraphics[width=\linewidth]{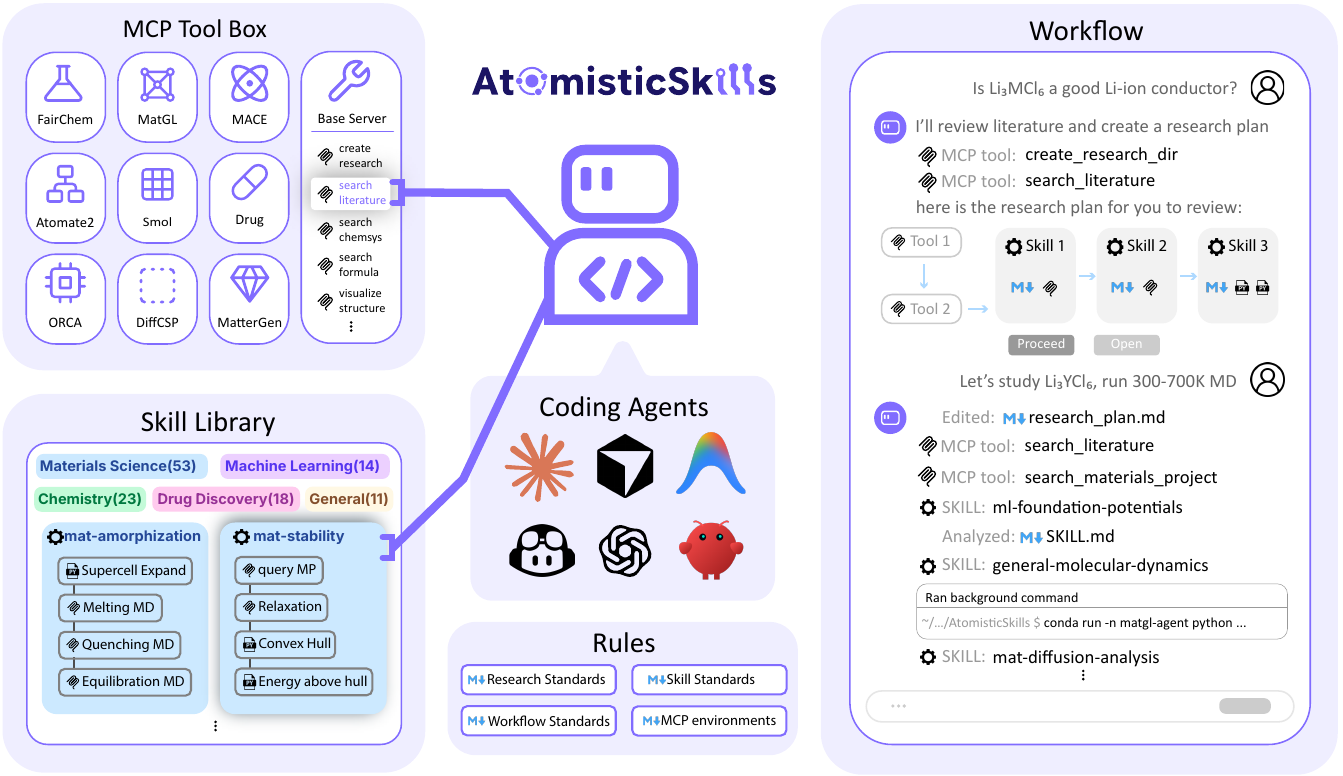}
\caption{\textbf{Architecture of AtomisticSkills}. The Model Context Protocol (MCP) toolbox provides fundamental research tools to the agent. Skill library contains mid-level, semi-flexible atomistic research skills that are each constructed with MCP tools, scripts, markdown files, and examples. Rules define research standards that the agent needs to follow. This hierarchical research infrastructure can be adopted by any general-purpose coding agent and enable its capability to perform atomistic research. In this way, complicated and versatile scientific research workflows can be hierarchically decomposed into skills and tools, and executed by coding agents under the supervision of human researchers.}
\label{fig:intro} 
\end{figure*}

The integration of computational workflows is central to modern chemical, materials, and biological research, enabling the efficient screening and design of novel batteries~\cite{Jun_2022_Lithium}, semiconductors~\cite{Walle_2004}, catalysts~\cite{Qiao_2011}, and drugs~\cite{halgrenMerckMolecularForce1996}. Despite its profound impact, the execution of these simulations involves significant operational overhead. Researchers must possess deep domain expertise not only in physics and chemistry but also in navigating fractured software ecosystems: parsing opaque parameters, handling runtime and convergence errors, and extracting data from complex, unstructured outputs. To automate these computational procedures, conventional high-throughput workflow frameworks have been developed~\cite{Ganose_2025_Atomate2, Curtarolo_2012_Aflow, uhrinWorkflowsAiiDAEngineering2021, jain_2015_fireworks, wang_2021_alkemie, weymuth_2024_scine}. These non-LLM-driven orchestrators are fundamentally hard-coded: while they provide exceptional reliability and reproducibility when executed under expected conditions, they remain non-expandable and require explicit programmatic definition for every new task. Consequently, they are primarily used for large-scale, systematic materials screening and deterministic data generation efforts, famously driving the creation of expansive databases like the Materials Project~\cite{Horton_2025_MP}. 

Recent advances in artificial intelligence (AI) and large language models (LLMs) are reshaping how scientific workflows are orchestrated, allowing systems to handle procedural steps that previously required human judgment and to adapt workflows dynamically in response to intermediate results. Indeed, numerous end-to-end architectures have emerged to tackle diverse challenges across different scientific domains. Hardware-integrated autonomous agents were reported focusing on closing the loop between digital planning and physical experimentation by orchestrating agentic reasoning, materials synthesis, and robotic characterizations~\cite{Boiko_2023_Autonomous, Song_2025_Multiagentdriven, Fei_2026_Agentic, Zhang_2025_Crest}. Moreover, knowledge retrieval and workflow navigation agents leverage structured databases, multi-disciplinary knowledge, and unbounded web search to dynamically read documentation, answer complex scientific examinations, generate hypotheses, and navigate pipelines with reduced hallucination rates~\cite{Chiang_2025_Llamp, Zou_2025_ElAgente, Gustin_2025_ElAgenteCuantico, Marwitz_2026}. In addition, agents also deployed \textit{in silico} simulation pipelines with various toolboxes to automate computational tasks, ranging from automated backend package deployment and multi-agent graph reasoning to orchestrating molecular dynamics (MD) simulations and first-principles calculations~\cite{Bran_2024_Augmenting, Ghafarollahi_2024_SciAgents, Hu_2025_Aitomia, Mitchener_2025_Kosmos, Zheng_2025_Large, Wang_2026_Deploymaster, Zhang_2025_Bohrium, Ruza_2026_reasoningtosimulation, Soleymanibrojeni_2026, Kumar_2026_ElAgenteSolido}.

Despite these successes, deploying end-to-end, domain-specific AI agents poses significant infrastructural bottlenecks. Most scientific agents were built as monolithic, standalone platforms atop orchestration frameworks such as LangChain~\cite{Chase_2022_LangChain}. Consequently, developers of these platforms are forced to reinvent fundamental software infrastructure: engineering system prompts for individual sub-agents, defining complex inter-agent communication boundaries, building custom user interfaces, and implementing persistent state tracking. In comparison to pre-LLM scientific computation management systems~\cite{Ganose_2025_Atomate2, Curtarolo_2012_Aflow, jain_2015_fireworks}, these agent platforms are difficult to transfer and deploy across different hardware environments due to additional dependency requirements on heavy agentic infrastructure and LLM calls. Furthermore, this monolithic design limits agility. It is challenging for end-to-end scientific agents to keep pace with the fast-evolving ecosystem of general-purpose agents, such as Claude Code~\cite{anthropicClaudeCode2025}, Aider~\cite{gauthierAider2023}, Codex~\cite{openaiCodex2025}, Gemini CLI~\cite{googleGeminiCli2025}, Cursor~\cite{anysphereCursor2023}, Google Antigravity~\cite{googleAntigravity2025}, and Windsurf~\cite{codeiumWindsurf2024}. These agents rapidly introduce novel LLM endpoints, refine orchestration graphs, and develop new standards for persistent memory and long-horizon tasks.

\begin{figure*}[t]
\centering
\includegraphics[width=\linewidth]{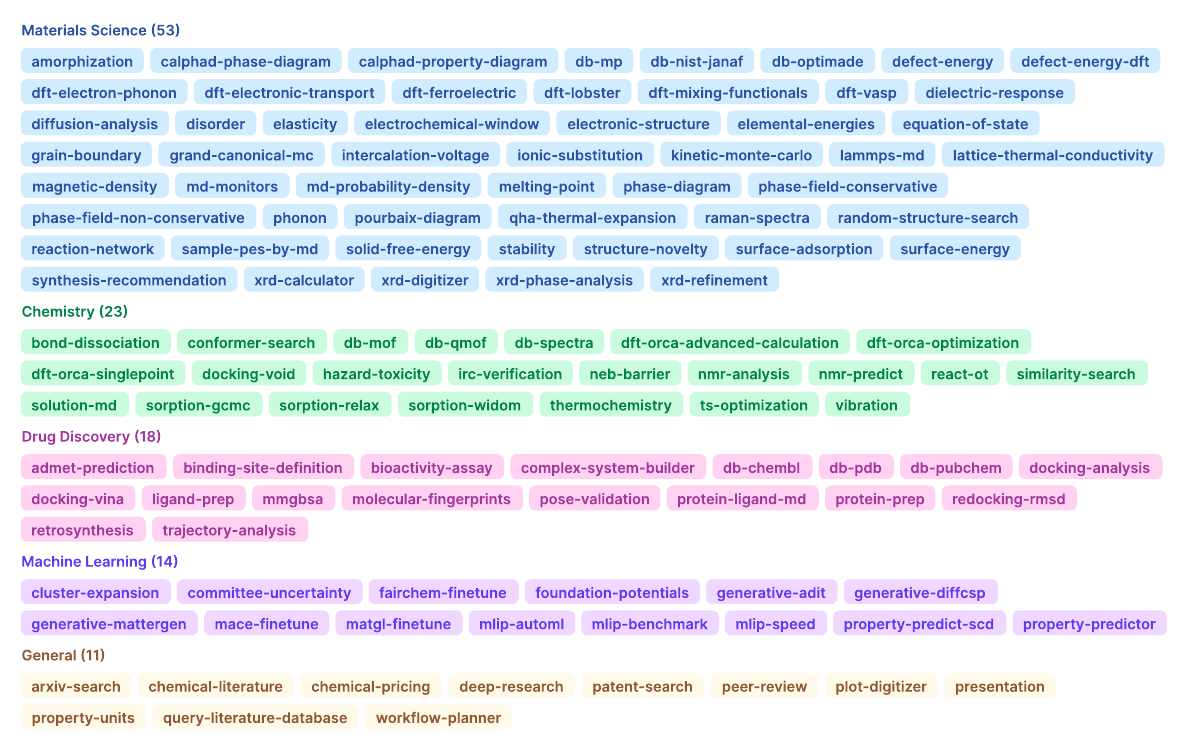}
\caption{\textbf{Skill library in AtomisticSkills.} AtomisticSkills provides comprehensive coverage of agentic research skills in materials science, chemistry, drug discovery, and machine learning. These skills serve as composable research units that can be built into full research workflows.}
\label{fig:skill_library} 
\end{figure*}

These limitations have motivated a complementary approach that builds on top of general-purpose coding agents rather than rebuilding their infrastructure: harness engineering~\cite{Zhou_2026_Externalization}. A harness is the non-model software infrastructure that wraps an LLM's reasoning loop and coordinates tool dispatch, context management, memory, and safety enforcement. Under the harness engineering approach, frontier coding agents serve as the substrate, and domain-specific capability is layered on as modular extensions. Recent work has shown that, for a fixed model, the choice of agent harness can be a primary determinant of end-to-end performance, capable of shifting outcomes by margins comparable to or exceeding model upgrades on agentic coding benchmarks~\cite{kapoor2025ai, kapoor2026holistic}. It is useful to disentangle two layers of this harness, defined here as the \textit{inner harness} and \textit{outer harness}. The inner harness, built by the agent provider, is the general-purpose machinery of the agent loop itself: system prompts, context compaction, built-in tool dispatch, and generic error handling (e.g., Claude Code or the Claude Agent SDK). The outer harness, built by the domain user, sits on top and supplies domain-specific procedural knowledge that informs the agent's reasoning before it acts. A highly beneficial way to shape the outer harness is via the construction of a library of domain-specific \textit{agent skills}, which were first introduced in this setting by Anthropic in late 2025~\cite{anthropicAgentSkills2025}. Skills are procedural documents that facilitate structured and controlled knowledge injection to the agent's context without the heavy pre-indexing required by conventional retrieval-augmented generation (RAG)~\cite{Lewis_2020_Retrievalaugmented}. This dynamic loading pattern is designed to address two well-documented LLM failure modes: performance degradation as context length grows~\cite{liu2024lost, hsieh2024ruler, modarressi2025nolima, hong2025context}, which persists even in modern frontier models with million-token windows, and the distractibility induced by irrelevant context even at short lengths~\cite{shi2023distract}. With agent skills, only a minimal \texttt{SKILL.md} documentation is loaded into the agent's context, as opposed to loading full multi-package documentation simultaneously.

The skill convention is very well-suited to the computational sciences, where workflows are typically highly procedural, environment-dependent, and tightly coupled to specialized tooling. Researchers can package transferable scientific procedures and toolsets as skill artifacts that work with any general-purpose coding agent. Each skill bundles natural-language procedures with actionable scripts and model context protocol (MCP)~\cite{Anthropic2024MCP} tools. The result is robust execution and cross-platform transferability without duplicating core infrastructural functionality like local file management and web search. Several projects have already moved in this direction, building curated modular skill libraries for domain-specific scientific work---among them ToolUniverse~\cite{Gao_2025_Democratizing}, Scientific Agent Skills~\cite{Anthropic_2026_Claude}, and MolClaw~\cite{Zhang_2026_MolClaw}. However, these efforts are almost exclusively concentrated in bioinformatics and clinical science. An open, systematic, and extensible skill framework dedicated to atomistic materials science and computational chemistry remains a critical open challenge.

To bridge this gap, we introduce AtomisticSkills, an open-source agentic research infrastructure designed for complex atomistic research. AtomisticSkills targets the outer harness for atomistic science. In this way, instead of reinventing the core agent loop itself, it layers atomistic research knowledge on top of any general-purpose coding agent. A critical challenge in deploying agentic research lies in resolving the inherent conflict between the deterministic reliability of traditional automation scripts and the unbounded, often hallucination-prone flexibility of free-form coding agents. AtomisticSkills addresses this by grounding unstructured LLM reasoning through a rigorous hierarchical decomposition of scientific knowledge into workflows, skills, and tools, as shown in Fig.~\ref{fig:intro}. 

Workflows represent comprehensive research campaigns targeted at high-level scientific objectives. As shown in the right panel of Fig.~\ref{fig:intro}, research directives such as ``Is \ch{Li3MCl6} a good Li-ion conductor?" are consolidated into modularized research plans by the coding agent with the aid of AtomisticSkills. Instead of uninformed execution, the agent first hypothesizes solutions, retrieves pertinent literature, and down-selects an appropriate sequence of skills to formulate a transparent plan. Following user review and approval, the agent then autonomously orchestrates the prolonged execution campaign.

Skills function as mid-level modular, semi-flexible research tutorials. Rather than standalone executables, skills consist of sequences of tools designed to solve specific scientific questions. For example, as detailed in the lower left of Fig.~\ref{fig:intro}, the \texttt{mat-amorphization} skill~\cite{Ishimaru_1997_amorphization} consists of a supercell expansion and three MD simulations, each corresponding to an MCP tool call. Similarly, the \texttt{mat-stability} skill involves an MCP-based database query~\cite{Horton_2025_MP}, ionic relaxation, and script-based construction and analysis of the convex hull phase diagram~\cite{Sun_2016_Thermodynamic}. Most skills in our library are implemented based on established methods documented in the literature and are explicitly paired with a markdown manual, executable scripts, and concrete example usages. To validate their reliability, execution results from these skills have been compared to literature-reported values whenever available. Additionally, to ensure complete scientific reproducibility, all hyperparameters and arguments invoked during a skill's execution are persistently logged into standardized configuration files alongside the generated outputs. 

The current skill library features more than 100 skills, spanning 53 skills in material science, 23 skills in chemistry, 18 skills in drug discovery, 14 skills in machine learning, and 11 general research skills. Fig.~\ref{fig:skill_library} catalogs the detailed skill coverages of current AtomisticSkills, which involve multi-disciplinary database access, simulation methods, machine learning models, experimental analysis, etc. To our best knowledge, AtomisticSkills presents the first agent skill infrastructure designed for general-purpose atomistic research. 

Tools constitute low-level, strictly structured operations exposed as Python functions, representing frequent, reusable fundamental research units that can be picked off-the-shelf to construct mid-level skills. They feature rigorous input/output type definitions and facilitate reliable, programmatic calls. Tools live in independent, separately operating environments (MCP servers) to avoid dependency conflicts. The top left panel of Fig.~\ref{fig:intro} shows a list of available MCP servers. In total, we provide more than 10 tool servers and 50 MCP tools. These include FairChem~\cite{Wood_2025_UMA}, MatGL~\cite{Ko_2025_MatGL}, and MACE~\cite{Batatia_2025_MACE}, which provide machine learning interatomic potential (MLIP) support for simulations; Atomate2~\cite{Ganose_2025_Atomate2}, which provides remote simulation management; Smol~\cite{Barroso_2022_smol}, which provides lattice thermodynamics analysis; materials and molecular generative models~\cite{Zeni_2025_Generative, Jiao_2024_DiffCSPPP}, and a drug server that provides drug discovery tools, etc. In addition to these universally reusable research tools implemented globally as MCP servers, AtomisticSkills also provides auxiliary tool environments that support specific and less reusable skill scripts that are not configured as persistently exposed MCP tools. Through these individual servers, the framework executes programmatic operations such as database retrieval, static energy calculations, and MDs via local or remote high-performance computing (HPC) tool calls. Because these tool payloads are strictly type-checked by the MCP protocol, deployed LLMs experience near-zero hallucination rates when configuring input parameters during these routine procedures. Their foundational reliability is continually guaranteed by unit testing (\texttt{pytest}).

In addition, rules form the guidelines and constraints of the agent's behavior. As shown in the lower center of Fig.~\ref{fig:intro}, rules establish formal research standards, such as mandating preliminary literature reviews, enforcing MLIP registry checks, managing simulation outputs, and defining the standards to create new tools, skills, and workflows. These agentic constraints ensure AtomisticSkills executes systematic and reproducible research.

This hierarchical decomposition empowers the agent to build complex scientific workflows dynamically from tested skills and tools, without sacrificing the deterministic reliability required for rigorous atomistic research. By leveraging the built-in code generation and localized tool-execution capability of general-purpose agents, AI scientists with AtomisticSkills can autonomously execute atomistic research workflows, spanning from ad-hoc literature retrieval to programmatic data analysis, while generating localized, human-readable state artifacts that permanently ensure step-by-step scientific reproducibility. In the next section, we will show several end-to-end case studies demonstrating how the framework seamlessly manages rigorous operational complexity across computational materials science, chemistry, biology, and experimental characterization analysis.

\section{Results}

\subsection{AtomisticSkills Coverage in Literature}

\begin{figure*}[th]
\centering
\includegraphics[width=\linewidth]{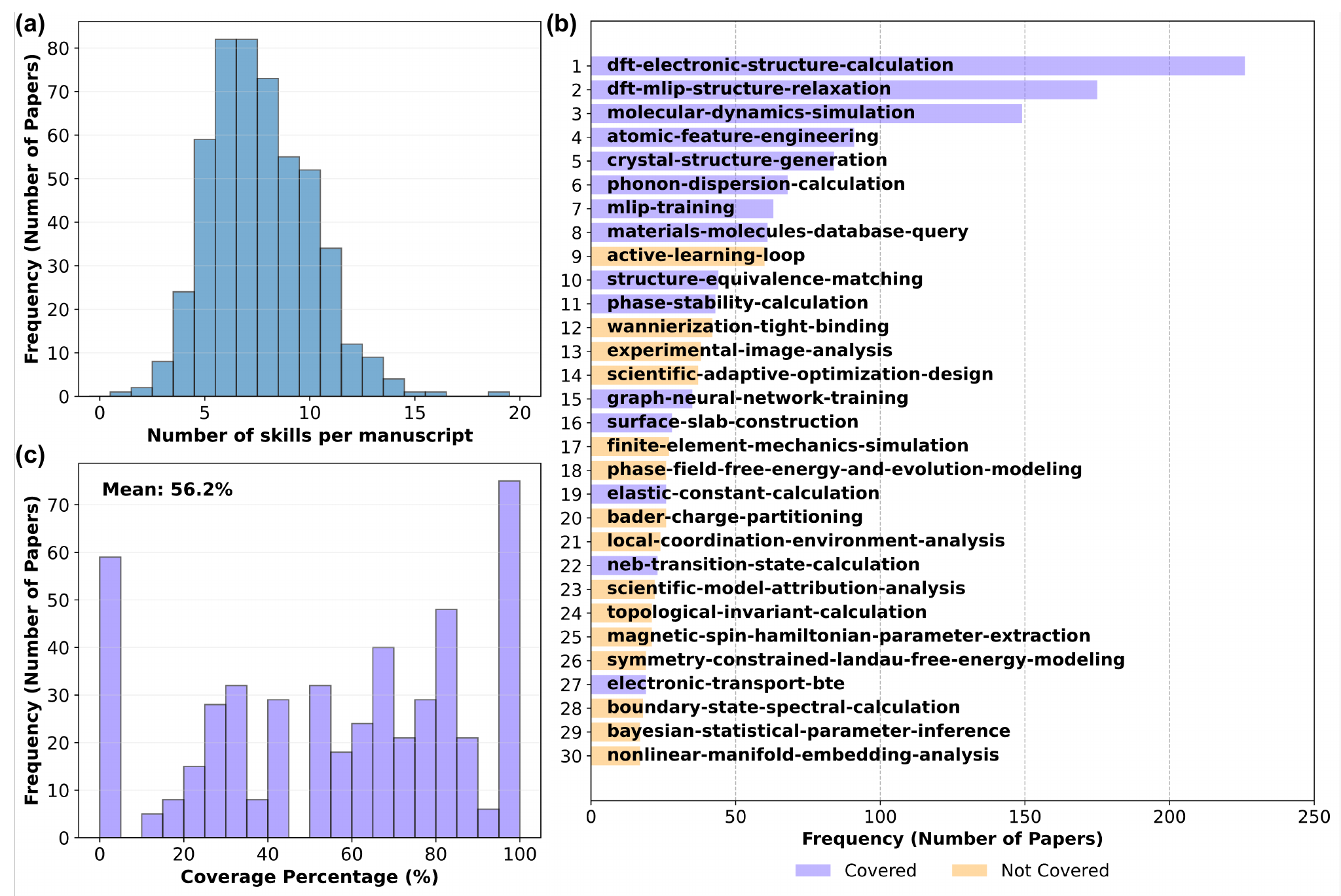}
\caption{\textbf{Coverage analysis of AtomisticSkills compared to literature} \textbf{(a)} Histogram of the number of skills used in each paper, text-mined from 500 published research articles from \textit{npj Computational Materials}. \textbf{(b)} The 30 most frequently used scientific skills. The histogram indicates their frequency appeared in literature, and the color scheme indicates whether the skill is covered by AtomisticSkill. \textbf{(c)} Histogram of the skill coverage percentage of AtomisticSkills in each paper. The result indicates that AtomisticSkills can fully cover around 15\% of computational materials journal articles. And on average, AtomisticSkills covers 56.2\% of computational materials skills usage.}
\label{fig:coverage} 
\end{figure*}

To benchmark the functional capabilities of AtomisticSkills, we conducted a systematic coverage analysis by text-mining and randomly sampling 500 papers published in \textit{npj Computational Materials}. By decomposing the computational workflows reported in each manuscript into their constituent skills (see Methods~\ref{sec:methods_skill_identification}), we systematically compared the literature-used skills against our library and quantitatively evaluated how effectively AtomisticSkills supports real-world materials research demands.

Fig.~\ref{fig:coverage}(a) illustrates the number of skills we identified from each article in \textit{npj Computational Materials}. The distribution indicates that typical computational studies rely on fewer than 10 interconnected skills. Crucially, evaluating these operations against our repository reveals a notably high degree of functional coverage. Fig.~\ref{fig:coverage}(b) highlights the 30 most frequently used skills found in the articles. The bar length indicates usage frequency, while the color denotes whether the literature-mined research operation is functionally supported by a skill in our library. The result shows that the frequently used skills in literature are mostly covered by AtomisticSkills. Finally, in Fig.~\ref{fig:coverage}(c), we plot the distribution of the percentage skill coverage in the 500 articles, which demonstrates that the framework provides reasonable support for the vast majority of computational materials science articles, with an average of 56.2\% coverage, and approximately 15\% of the evaluated literature reaching 100\% operational mapping. Notably, there are around 12\% of journal articles with 0 coverage. These missing coverages mainly correspond to articles with novel algorithmic developments~\cite{Na_2025_npj_method}, experimental methods~\cite{Ueno_2018_npj_experiment}, and special domains that require less frequent computational skills~\cite{Liu_2019_npj_special, Zhao_2018_npj_special}.

Furthermore, to validate the framework's broader applicability in computational chemistry, we conducted a similar skill coverage analysis on 500 open-access articles from \textit{The Journal of Physical Chemistry Letters}. For computational drug discovery, we collected 500 open-access articles filtered with keywords from PubMed and arXiv. The results show that AtomisticSkills have 44.9\% coverage of computational chemistry journal articles and 62.4\% coverage of computational drug discovery articles. These analyses are detailed in Supplementary Figs. S1 and S2.

\subsection{Generative design of Li-ion solid conductor}

\begin{figure*}[t]
\centering
\includegraphics[width=\linewidth]{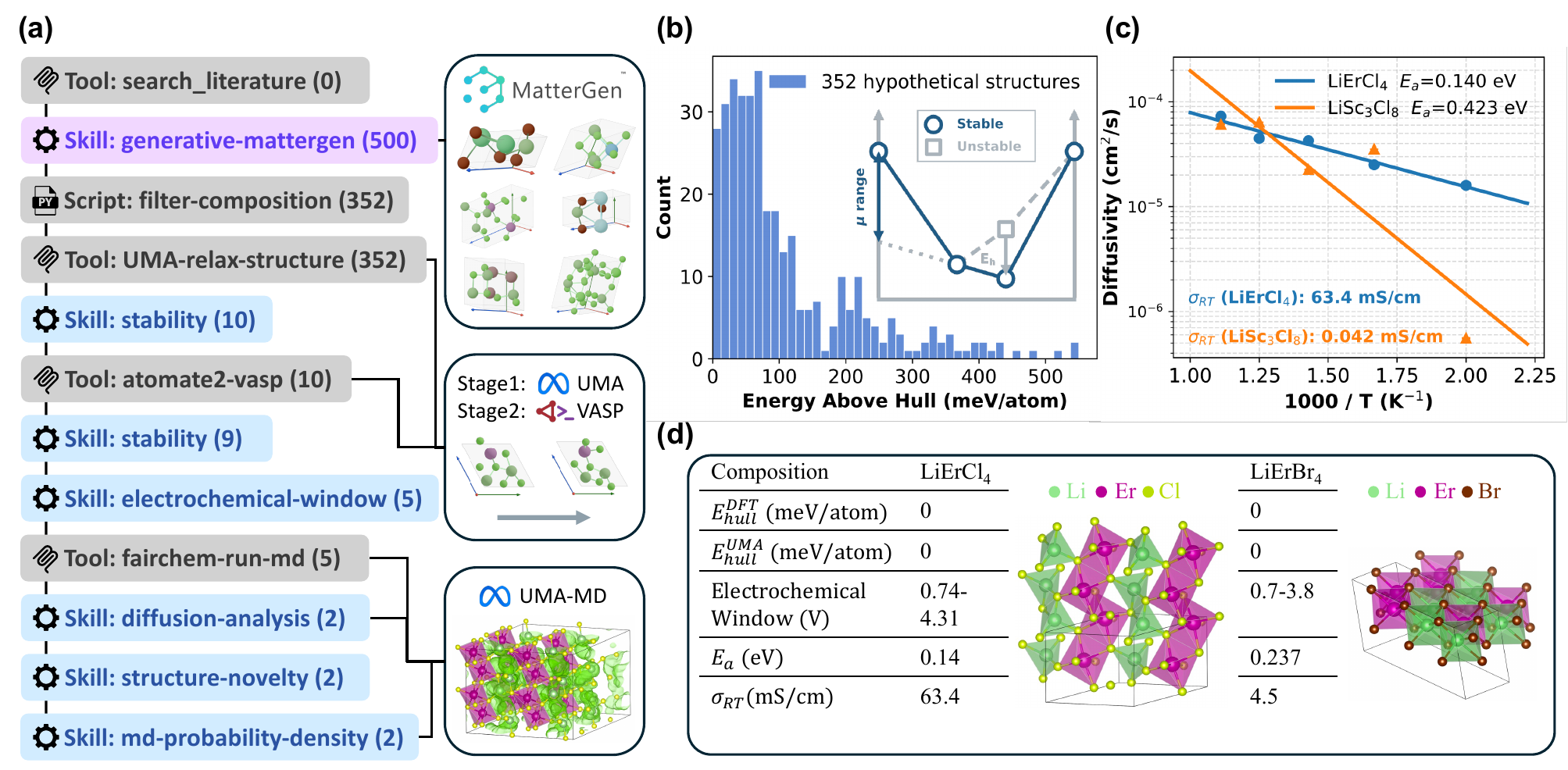}
\caption{\textbf{Generative design of Li-ion solid state electrolyte.} \textbf{(a)} Agentic workflow that involves candidate materials generation with MatterGen, screening of thermodynamic stability and electrochemical stability with foundation MLIPs and DFT, and diffusion analysis through MD simulation. The number of candidates after each step is indicated in parentheses. \textbf{(b)} Distribution of UMA-calculated energy above hull of 352 MatterGen-generated halide-based Li-ion SSE candidates. The inset plot illustrates how $E_h$ and $\mu$ range are determined from constructing a convex hull phase diagram. \textbf{(c)} Arrhenius plot and extracted room-temperature Li-ion conductivity of two candidate materials from the MD simulation with \texttt{-s-1p2-omat}. \textbf{(d)} Structures and calculated properties of two top halide Li-ion SSEs.}
\label{fig:halide} 
\end{figure*}

Identifying novel superionic solid-state electrolytes (SSEs) is pivotal to advancing safe, high-energy-density solid-state batteries. Discovering materials that are simultaneously thermodynamically stable and highly ionically conductive involves exploring a vastly uncharted chemical space. To address this, we demonstrate an end-to-end AtomisticSkills agentic workflow to discover halide-based Li-ion SSEs, illustrated in Fig.~\ref{fig:halide}(a). This high-throughput computational screening pipeline integrates generative models, foundation MLIPs (FPs), and DFT. The pool of material candidates is progressively narrowed after each operational step, with the remaining subset sizes denoted in parentheses. In Fig.~\ref{fig:halide}(a) and all subsequent workflow charts, box colors indicate skill categories, consistent with the schema in Fig.~\ref{fig:skill_library}. For clarity, category prefixes such as 'ml-' and 'mat-' are omitted in the plots but included in the manuscript.

With the \texttt{ml-generative-mattergen}~\cite{Zeni_2025_Generative} skill, the agent first generated 500 hypothetical Li-containing halide structures across 10 ternary chemical spaces: Li-Er-Br, Li-Er-Cl, Li-Hf-Cl, Li-In-Br, Li-In-Cl, Li-Sc-Br, Li-Sc-Cl, Li-Y-Br, Li-Y-Cl, and Li-Zr-Cl. Then, a composition filter was applied to remove all generated materials that were not charge-neutral according to common elemental valence states. These candidates were subsequently filtered using the \texttt{mat-stability} and \texttt{mat-electrochemical-window} skills, which evaluated thermodynamic stability ($E_{\text{hull}}$)~\cite{Sun_2016_Thermodynamic} and the intrinsic electrochemical window (ECW)~\cite{Zhu_2015_Origin} against \texttt{UMA-s-1p2-omat}~\cite{Wood_2025_UMA} and DFT constructed convex hulls. Specifically, only structures with $E_{\text{hull}} = 0$ meV atom$^{-1}$ and ECW $< 2$ V~\cite{Zhu_2015_Origin} were retained.

Fig.~\ref{fig:halide}(b) shows the distribution of energies above the hull ($E_{\text{hull}}$) calculated by the \texttt{UMA-s-1p2-omat}~\cite{Wood_2025_UMA}, where candidates with lower $E_{\text{hull}}$ are thermodynamically more stable. The inset schematic plot illustrates how the $E_{\text{hull}}$ and stable chemical potential ($\mu$) ranges are calculated from the 0 K computational convex hull phase diagram~\cite{Sun_2016_Thermodynamic}. The top candidates then underwent MCP-tool-triggered multi-temperature MD simulations with \texttt{UMA-s-1p2-omat}~\cite{Wood_2025_UMA} and the \texttt{mat-diffusion-analysis} skill to extract ionic conductivities and activation barriers. Top candidates were filtered with an extrapolated room-temperature ionic conductivity $< 1$ mS cm$^{-1}$.

The Arrhenius plot in Fig.~\ref{fig:halide}(c) demonstrates the extraction of activation energy ($E_a$) and room-temperature Li-ion conductivity ($\sigma_{\text{RT}}$) for the simulated candidates. The ionically conductive SSE candidate (green) exhibits lower $E_a$ and higher $\sigma_{\text{RT}}$ compared to the poorly conducting SSE candidate (red). Finally, Fig.~\ref{fig:halide}(d) displays the structural configurations and quantitative properties of the top two halide Li-ion SSEs discovered in this exploration. Notably, the agent discovered two previously unreported SSE compounds, \ch{LiErCl4} and \ch{LiErBr4}, which are thermodynamically stable according to the DFT-constructed convex hull phase diagram and exhibit high predicted conductivities of 63.4 and 4.5 mS cm$^{-1}$ with low activation barriers of 0.14 and 0.237 eV, respectively. 

It should be noted that the simulation hyperparameters used in this workflow, such as MD simulation lengths ($\sim$20 to 100 ps based on early-convergence stopping) and screening thresholds, were suggested by the agent and reviewed by human researchers during the research planning (see Fig.~\ref{fig:intro}) to align with reported SSE high-throughput screening studies~\cite{Jun_2022_Lithium, Chen_2024}. As a result, the simulated conductivities serve strictly as literature-standard computational screening metrics rather than rigorously converged values targeted for direct experimental validation. Nevertheless, this abbreviated workflow demonstrates the capacity to construct fully agentic, rigorous materials discovery pipelines with AtomisticSkills. Ultimately, by seamlessly chaining machine learning algorithms, thermodynamic stability analyses, and advanced simulation methods, the modularized skill design enables the potential for fully autonomous materials discovery at vast scales.

\subsection{Computational screening of MOF for \ch{CO2} capture}

\begin{figure*}[t]
\centering
\includegraphics[width=\linewidth]{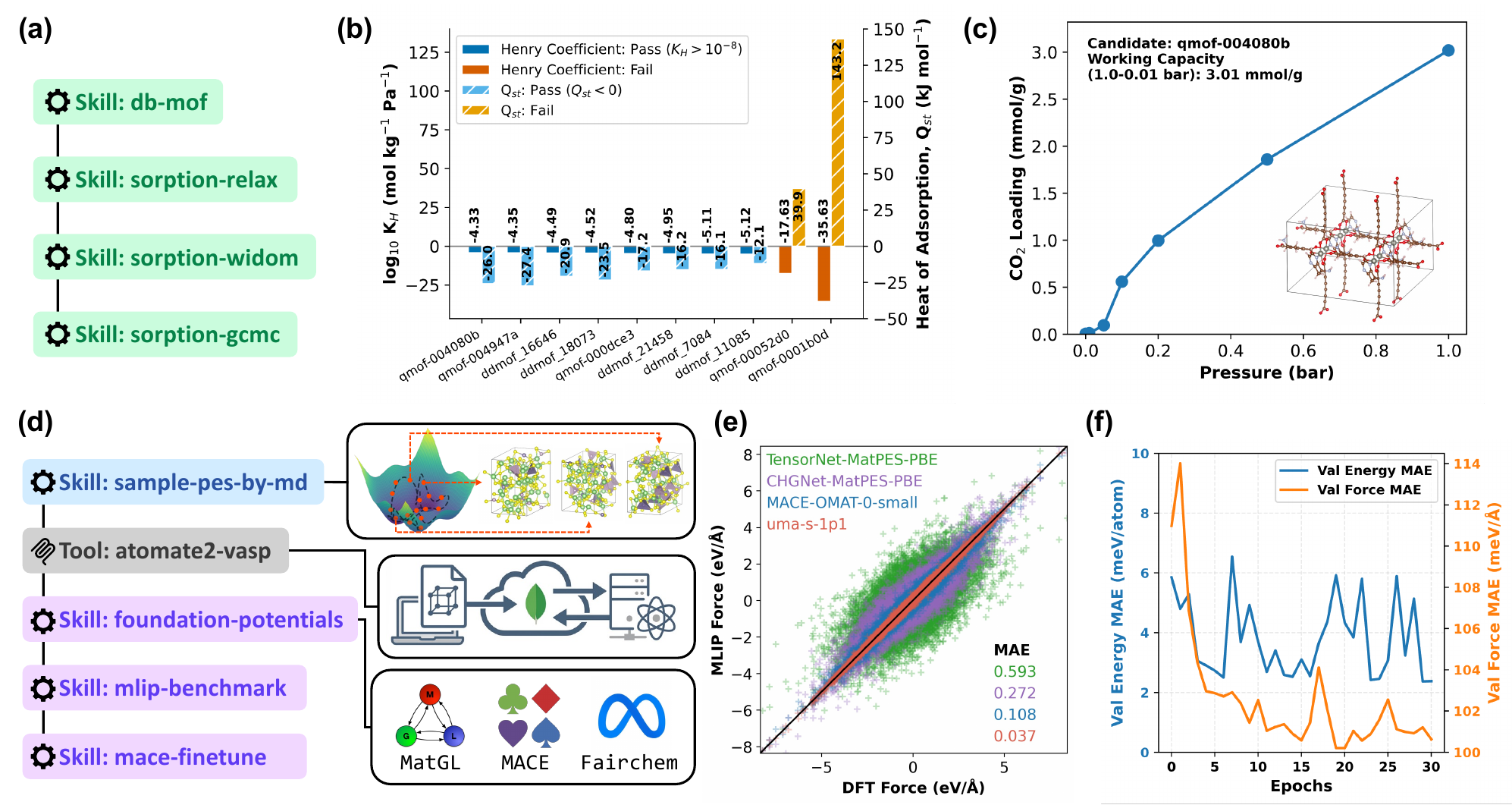}
\caption{\textbf{Agentic screening of MOFs and MLIP AutoML workflows.} \textbf{(a)} Agentic computational screening workflow for MOF-based \ch{CO2} capture. \textbf{(b)} Calculated Henry coefficient and \ch{CO2} heat of adsorption at 298~K, for 10 MOF candidates queried from QMOF and ARC-MOF databases. \textbf{(c)} \ch{CO2} adsorption isotherm calculated at 298~K via GCMC simulation leveraging \texttt{UMA-s-1p2-odac}. \textbf{(d)} Agentic MLIP workflow composed of automatic sampling of potential energy surface, DFT calculations on remote HPC, selection of FPs, benchmarking, and fine-tuning. \textbf{(e)} Parity plot and force MAE for 4 FPs compared to DFT ground-truth labels on the \ch{Li10GeP2S12} dataset. \textbf{(f)} Agentic fine-tuning MAE history of MACE-OMAT-0-small on the sampled dataset.}
\label{fig:mof_mlip} 
\end{figure*}

Metal-Organic Frameworks (MOFs) exhibit profound structural and chemical tunability, making them premier candidates for both point-source (industrial flue gas) and dilute source (Direct Air Capture (DAC)) \ch{CO2} capture technologies~\cite{majumdar2021diversifying,park2024inverse}. While end-to-end agentic MOF pipelines have recently been demonstrated in literature~\cite{Zheng_2025_Large, Kang_2024_Chatmof}, we demonstrate that similar agentic discovery pipelines can be constructed in AtomisticSkills without explicit predefined orchestration graphs. Discovering optimal MOFs requires screening massive structural databases to identify optimal pore geometries and chemical environments that maximize selective gas uptake. To address this MOF screening challenge, the AtomisticSkills agent planned a hierarchical screening workflow shown in Fig.~\ref{fig:mof_mlip}(a). It first deployed the \texttt{chem-db-mof} skill to autonomously retrieve candidate architectures from curated databases, specifically QMOF~\cite{rosen2021machine} and ARC-MOF~\cite{burner2023arc}, then used the \texttt{chem-sorption-relax} and \texttt{chem-sorption-widom} skills to perform rapid Widom insertions~\cite{Widom_1963_Some} on these queried frameworks using \texttt{UMA-s-1p2-odac}. The high-capacity candidates were subsequently passed into the \texttt{chem-sorption-gcmc} skill for rigorous, multi-pressure grand canonical Monte Carlo (GCMC) isotherm evaluations.

Fig.~\ref{fig:mof_mlip}(b) plots the Henry coefficients ($K_H$) alongside the \ch{CO2} heat of adsorption at infinite-dilution ($Q_{st}$) for ten MOFs queried from the QMOF and ARC-MOF libraries. Here, heuristic filtering thresholds indicative of thermodynamically viable physisorbents ($K_{H} > 10^{-8}$ mol kg$^{-1}$ Pa$^{-1}$ and $Q_{st} < 0$ kJ mol$^{-1}$) are used to identify promising structures for DAC. Subsequently, Fig.~\ref{fig:mof_mlip}(c) shows the GCMC-calculated \ch{CO2} adsorption isotherm for the top candidate \texttt{qmof-004080b}, with 50,000 GCMC steps used at each pressure. The starting point of the isotherm, also known as the Henry's regime, represents ultra-dilute atmospheric \ch{CO2} relevant for DAC purposes~\cite{Kumar_2015_DAC}. In comparison, the 1 bar region caters more to post-combustion applications, such as flue gas point-source capture~\cite{Samanta_2011_Post}. 

Notably, the candidate \texttt{qmof-004080b} exhibits a calculated Henry coefficient of $4.64 \times 10^{-5}$ mol kg$^{-1}$ Pa$^{-1}$ ($\log_{10} K_H \approx -4.33$) and a heat of adsorption ($Q_{st}$) of $-$26.0 kJ mol$^{-1}$, hitting the ``sweet spot'' of modern \ch{CO2} physisorbents~\cite{Jiang_2023_Sorption} by balancing robust uptake with manageable regeneration energy. While its simulated \ch{CO2} working capacity of $\sim 3.0$ mmol g$^{-1}$ ($1.0$--$0.01$ bar) demonstrates promising performance for post-combustion capture, the GCMC result at the ultra-dilute 0.0004 bar regime reveals a negligible loading, indicating \texttt{qmof-004080b} is not ideal for DAC application. This screening workflow on a small candidate pool demonstrates the capability of AtomisticSkills to bridge structural curation to sorption evaluation. To acquire additional high-fidelity results, the agentic workflow can be seamlessly extended to assess humidity stability via MD or adsorption kinetics via DFT.

\subsection{Agentic MLIP benchmark and fine-tuning}

Foundation MLIPs, also known as FPs, have demonstrated broad applicability and generalizability across chemical spaces \cite{Chen_2022_M3GNet, Jacobs_2025, Yuan_2026, Deng_2023_CHGNet, Batatia_2025_MACE, Wood_2025_UMA, Kaplan_2025_MatPES, Perez_2026_SCD}. However, they can suffer from degraded accuracy when extrapolating to out-of-distribution atomistic configurations, such as those encountered during high-temperature MD~\cite{Deng_2025_Systematic, Fu_2022_Forces}. To ensure their reliability, systematic benchmarking~\cite{Riebesell_2025_MBD} and data-efficient fine-tuning are often required~\cite{Qi_2024_sampling}. However, divergent syntaxes, incompatible data pipeline formats, and conflicting environment dependencies across MLIP architectures pose a technical barrier for practitioners. These challenges motivate agentic orchestrations of complex machine learning workflows entirely via natural language~\cite{Lahouari_2025_MLIPagent}.

In this section, we showcase a fully agentic MLIP benchmarking and fine-tuning workflow focused on the superionic SSE \ch{Li10GeP2S12}~\cite{Ong_2012_LGPS}. Fig.~\ref{fig:mof_mlip}(d) illustrates this automated machine learning (AutoML) cycle. The agent initially sampled 200 structures from the \ch{Li10GeP2S12} potential energy surface (PES) by utilizing the \texttt{mat-sample-pes-by-md} skill~\cite{Qi_2024_sampling}. Subsequently, remote HPC-DFT jobs with MatPES-VASP settings~\cite{Kaplan_2025_MatPES} were submitted via the \texttt{atomate2-vasp} MCP tool to acquire high-fidelity energy and force labels for the sampled structures. The \texttt{ml-foundation-potentials} skill guided the agent to identify the most relevant FPs for the \ch{Li10GeP2S12} system, which was followed by the execution of the \texttt{ml-mlip-benchmark} skill. Here, the agent benchmarked the energy and force prediction accuracy of four selected FPs: TensorNet~\cite{Ko_2025_MatGL}, CHGNet~\cite{Deng_2023_CHGNet}, MACE~\cite{Batatia_2025_MACE}, and UMA~\cite{Wood_2025_UMA}. The resulting force parity plots and force mean absolute errors (MAEs) are shown in Fig.~\ref{fig:mof_mlip}(e). AtomisticSkills further facilitated the agentic fine-tuning of these models. Following the instructions provided by the \texttt{ml-mace-finetune} skill, the agent autonomously fine-tuned the MACE-OMAT-0-small architecture using the prepared structures and labels. Fig.~\ref{fig:mof_mlip}(f) charts the energy and force validation MAE history during this agent-triggered fine-tuning, demonstrating a gradual decrease in validation MAEs over a brief 30-epoch training run.

These findings demonstrate that atomistic ML models, such as FPs, can be efficiently benchmarked and fine-tuned using the AtomisticSkills framework. Through the autonomous management of training data preparation, remote HPC job submission, and complex machine learning pipelines, AtomisticSkills significantly reduces the learning curve and operational complexity of advanced scientific computing infrastructures.

\subsection{Structure-based high-throughput virtual screening for drug discovery}

\begin{figure*}[ht]
\centering
\includegraphics[width=\linewidth]{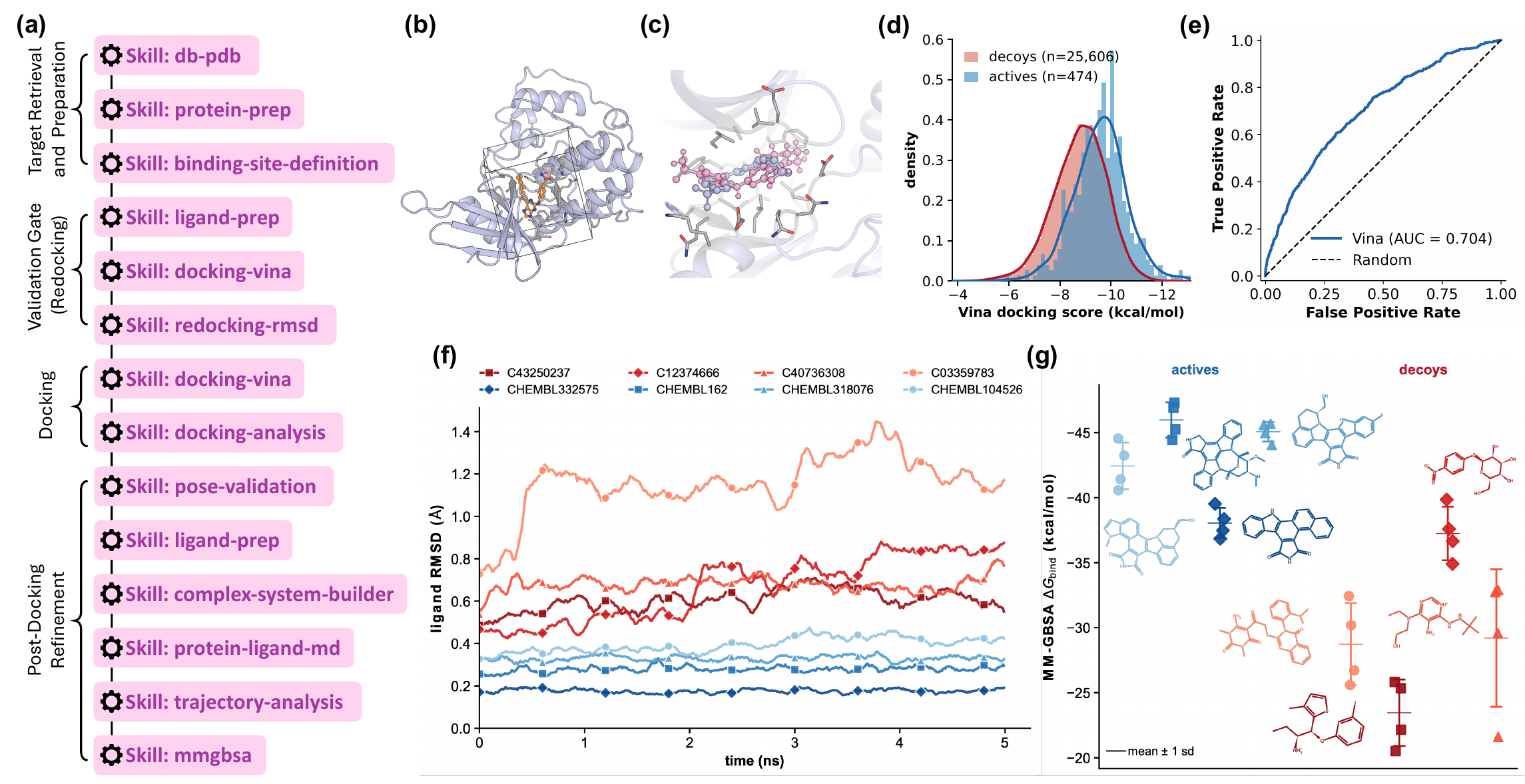}
\caption{\textbf{Agentic high-throughput virtual screening for drug discovery} \textbf{(a)} Computational docking pipeline as executed by the agent on the DUD-E CDK2 benchmark, separated into four main stages: target retrieval and preparation, redocking known actives for docking protocol validation, the primary docking run, and post-docking refinement. \textbf{(b)} Prepared CDK2 receptor (PDB: 1H1S) with autodefined docking box (\texttt{drug-binding-site-definition}) around the ATP-binding site; co-crystal ligand shown in orange. \textbf{(c)} Top-ranked Vina pose for the rank-1 active CHEMBL104526 in the CDK2 ATP-binding site. \textbf{(d)} Best-pose Vina docking-score distributions for the 474 known actives (blue) and 25{,}606 decoys (red) of the DUD-E CDK2 library, shown as kernel density estimates overlaid on normalized histograms. \textbf{(e)} ROC curve for Vina-based active/decoy classification on the DUD-E CDK2 benchmark (AUC = 0.70). Ligand heavy-atom RMSD over 5 ns of explicit-water MD after protein-C$\alpha$ alignment for four strong-Vina actives (ranks 1, 2, 3, 5; blue) and four weak-Vina decoys (ranks 20{,}000, 22{,}413, 24{,}022, 25{,}629; red). Each trace is the mean across four independent replicates with different integrator seeds, with a 110 ps centered rolling mean applied for visual smoothing. Marker shape identifies the compound and is shared with panel (g). \textbf{(g)} Per-replicate single-trajectory MM-GBSA binding free energy ($\Delta G_{\mathrm{bind}}$; GBn2 implicit solvent, no entropy correction) for the same eight compounds, with mean $\pm$ 1 s.d.\ error bars across the four MD replicates; stronger binding is plotted upward. Marker shape matches panel (f). The replicate means show clear class separation (active mean $-42.9 \pm 3.6$ vs.\ decoy mean $-29.7 \pm 5.7$ kcal mol$^{-1}$; Welch's $t$-test $p = 0.011$).}
\label{fig:docking} 
\end{figure*}

Structure-based virtual screening (SBVS) is commonly used to prioritize small molecules against a protein target before experimental assay~\cite{liontaStructureBasedVirtualScreening2014, lyuUltralargeLibraryDocking2019}, and in practice, each campaign requires an expert to orchestrate a chain of specialized tools (receptor preparation, docking, scoring, pose refinement) with judgment calls at every stage that make simple, deterministic automation difficult~\cite{benderPracticalGuideLargescale2021}. We tested whether an AI agent using AtomisticSkills can drive this workflow autonomously by running the DUD-E CDK2 benchmark (474 actives and 27,850 property-matched decoys)~\cite{mysingerDirectoryUsefulDecoys2012} end-to-end. From a single user prompt, the agent retrieved the holo CDK2 structure (PDB: 1H1S~\cite{bermanProteinDataBank2000, daviesStructurebasedDesignPotent2002}) using \texttt{drug-db-pdb}, repaired and protonated the receptor with \texttt{drug-protein-prep}, and defined the docking box from the co-crystal ligand via \texttt{drug-binding-site-definition}. It then generated 3D conformers and AutoDock-readable PDBQT files for the full library with \texttt{drug-ligand-prep} (RDKit ETKDG~\cite{rinikerBetterInformedDistance2015} / MMFF94~\cite{halgrenMerckMolecularForce1996} / Meeko~\cite{santosmartinsMeekoMoleculeParametrization2025}), ran AutoDock Vina~\cite{trottAutoDockVinaImproving2010} at exhaustiveness 32 through \texttt{drug-docking-vina}, and combined the resulting affinities into a ranked hit list. To determine whether the docked poses remain stable under explicit-solvent MDs, the agent built parameterized protein-ligand systems with \texttt{drug-complex-system-builder} (OpenFF Sage 2.2.0~\cite{boothroydDevelopmentBenchmarkingOpen2023} + Amber ff14SB~\cite{maierFf14SBImprovingAccuracy2015} + TIP3P~\cite{jorgensenComparisonSimplePotential1983}), ran 5 ns of explicit-water MD per compound via \texttt{drug-protein-ligand-md} (OpenMM~\cite{eastmanOpenMM8Molecular2024}, using a 4 fs time step enabled by hydrogen mass repartitioning~\cite{hopkinsLongTimeStepMolecularDynamics2015}), computed ligand heavy-atom RMSD trajectories with \texttt{drug-trajectory-analysis}, and estimated relative binding free energies with single-trajectory MM-GBSA~\cite{genhedenMMPBSAMM2015} in GBn2 implicit solvent~\cite{nguyenImprovedGeneralizedBorn2013} via \texttt{drug-mmgbsa}.

The screen is summarized in Fig.~\ref{fig:docking}. Panel (a) shows the skill-call sequence executed by the agent. Panel (b) shows the prepared receptor with the autodefined docking box, and panel (c) shows the top-ranked docked pose in the CDK2 ATP-binding site. Panel (d) shows the Vina score distributions for the 474 actives and 25{,}606 decoys separately, where the actives' density is shifted toward stronger predicted binding by $\sim$1 kcal mol$^{-1}$ but with substantial overlap, the visual basis for the AUC = 0.70 in panel (e). Panel (e) shows the receiver operating characteristic (ROC) curve and reports two standard SBVS metrics. The area under the curve (AUC) is the probability that a randomly drawn active is ranked above a randomly drawn decoy, where 0.5 corresponds to random ranking and 1.0 to perfect; here, AUC = 0.70. The enrichment factor at the top 1\% of the ranked list (EF${1\%}$) is the fold-enrichment of actives in the top 1\% relative to the population active rate; here, EF${1\%}$ = 7.0. Both values fall within the range typically reported for Vina on DUD-E targets~\cite{eberhardtAutoDockVina1202021}. To probe pose stability across the Vina ranking, the agent ran four independent 5 ns explicit-water MD replicates for each of four strong-Vina actives (ranks 1, 2, 3, 5) and four weak-Vina decoys (ranks 20{,}000, 22{,}413, 24{,}022, 25{,}629). Panel (f) shows the replicate-mean ligand heavy-atom RMSD: the actives all stabilize below 0.5 \AA, while the decoys spread between 0.5 and 1.4 \AA with one (C03359783) drifting upward over the trajectory. The MD step thus contributes pose-stability information that is not present in the docking score alone. Panel (g) shows the per-replicate MM-GBSA binding free energy ($\Delta G_{\mathrm{bind}}$) for the same eight compounds, with each compound's four replicate means displayed individually and error bars showing $\pm$1 s.d.\ across replicates. The class means are well-separated (actives $-42.9 \pm 3.6$ vs decoys $-29.7 \pm 5.7$ kcal mol$^{-1}$; Welch's $t$-test $p = 0.011$), thereby corroborating the RMSD-based pose-stability analysis with an independent energetic signal.

These results show how AtomisticSkills can autonomously drive a complete SBVS funnel, going from a raw PDB file to ranked hits with downstream stability filtering, which was able to successfully separate actives from decoys at a statistically significant level. The AUC and early-enrichment values are consistent with the published Vina literature, and the rank-stratified RMSD/MM-GBSA result demonstrates that the agent can compose multiple skills to complete post-docking refinement and ranking procedure from structural and energetic signals.

\subsection{Multimodal analysis of XRD patterns}

\begin{figure*}[ht]
\centering
\includegraphics[width=\linewidth]{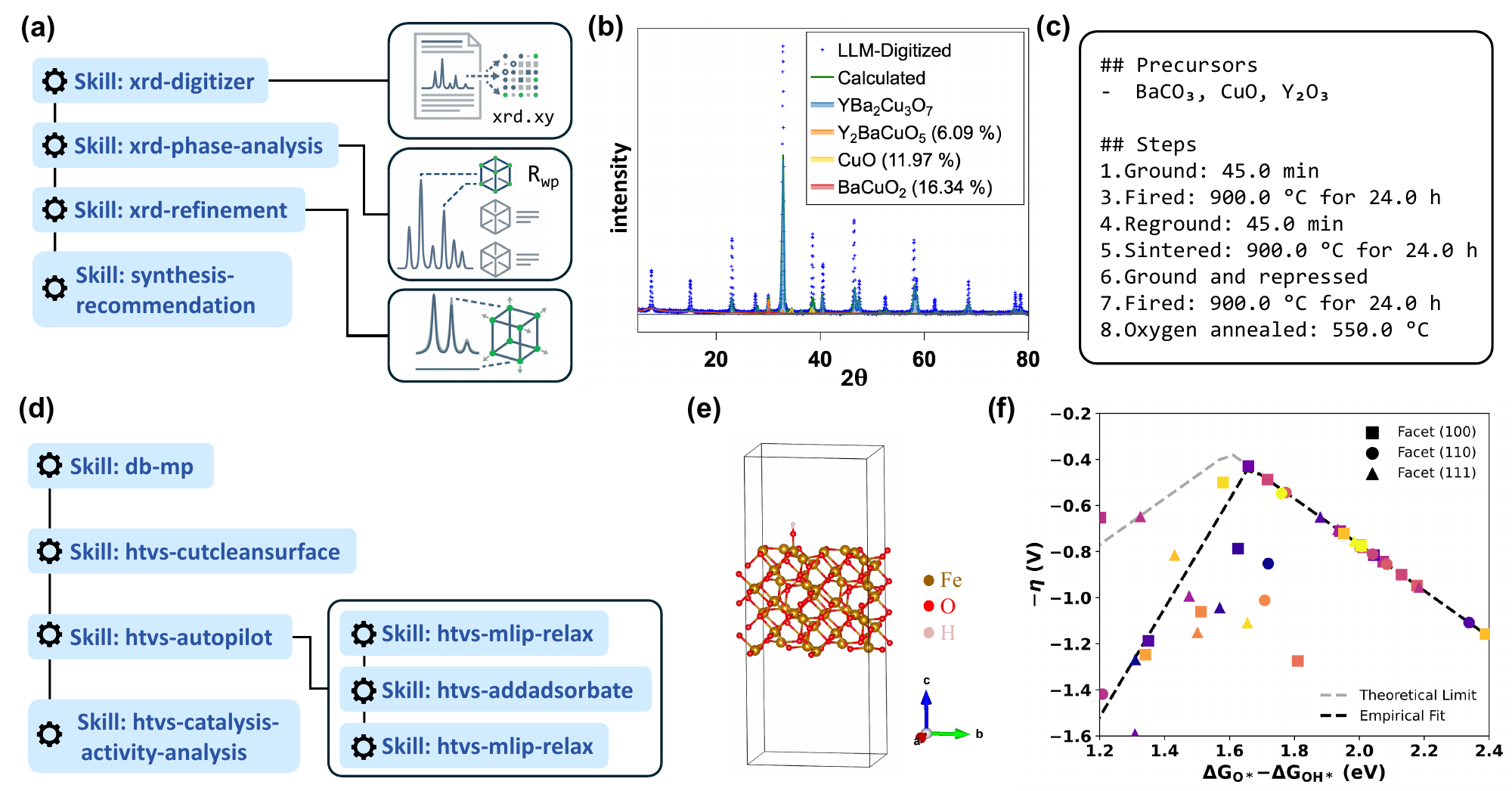}
\caption{\textbf{Agentic Analysis of XRD spectrum and screening of catalytic surfaces.} \textbf{(a)} AtomisticSkills workflow for analyzing experimental XRD spectrum. The user first sends a screenshot of an XRD pattern from an experimental characterization journal article to the agent. The \texttt{xrd-digitizer} skill then invokes the vision language model backend to digitize the screenshot into curve data, and passes the data into \texttt{xrd-phase-analysis} and \texttt{xrd-refinement} skills. Finally, \texttt{synthesis-recommendation} skill is called to propose a synthesis recipe for the target material. \textbf{(b)} The agentic XRD refinement result shows material compositions matching literature results. \textbf{(c)} Synthesis recommendation with precursors and synthesis steps. \textbf{(d)} Agentic workflow for screening Fe-oxide surfaces for oxygen evolution reaction (OER) with \texttt{htvs} remote job manager. \textbf{(e)} Representative atomic configuration of OH* on \ch{Fe3O4} (100) surface with Materials Project id \texttt{mp-19306}. \textbf{(f)} Volcano plot of MLIP prediction illustrating the OER overpotential as a function of the binding energy descriptor $\Delta G_{\rm O*} - \Delta G_{\rm OH*}$, with colored markers denoting distinct bulk polymorphs. The black dashed lines represent the empirical fit of the sampled data, while the grey dashed lines represent the universal theoretical limits~\cite{man2011universality}.}
\label{fig:xrd_catalysis} 
\end{figure*}

In contrast to highly standardized simulation outputs, the automated analysis of experimental characterization data presents a profound challenge for both human researchers and emerging AI scientists. This bottleneck is primarily driven by two factors: first, raw data formats are often dictated by proprietary characterization hardware and differ substantially between facilities; second, existing data within the broader literature is frequently presented in un-digitized figures rather than raw numerical arrays. Extracting actionable representations from these graphical data is a critical challenge in data-driven analysis. Overcoming this requires a truly multimodal approach capable of transforming unstructured visual data into machine-readable numerical arrays. While most agentic scientific frameworks restrict their operational scope to purely textual environments, AtomisticSkills natively integrates multimodal capabilities through its integration in general-purpose agents. This enables assimilation of the diverse, non-standardized visual data formats that are inherent to scientific research. 

To demonstrate this, we used AtomisticSkills to analyze an X-ray diffraction pattern (XRD) reported in a recent study~\cite{Turayev_2025_Xray}. Fig.~\ref{fig:xrd_catalysis}(a) shows the complete workflow. We began by sending the AtomisticSkills agent a screenshot of Fig.~2 from~\citet{Turayev_2025_Xray}, which provides an experimental XRD profile of the author-synthesized orthorhombic YBa$_{2}$Cu$_{3}$O$_{7-\delta}$ (YBCO-123) high-temperature superconductor. AtomisticSkills agent used the \texttt{mat-xrd-digitizer} skill, invoking integral Vision-Language Models (VLMs) to process the screenshot and digitize localized $2\theta$ peak intensities into a numerical array \texttt{xrd.xy}. The agent then translated this array into the data-driven automated Rietveld analysis (Dara)~\cite{Fei_2026_Dara} by routing the extracted arrays through the \texttt{mat-xrd-phase-analysis} in the Y-Ba-Cu-O chemical space, which autonomously and correctly identified the YBa$_{2}$Cu$_{3}$O$_{7-\delta}$ structure as the globally dominant crystalline phase.

Armed with this compositional hypothesis, the agent routed the data to the \texttt{mat-xrd-refinement} skill to perform a multiphase Rietveld refinement encompassing the primary YBa$_{2}$Cu$_{3}$O$_{7-\delta}$ phase alongside secondary phases \ch{Y2BaCuO5}, \ch{CuO}, and \ch{BaCuO2}. Fig.~\ref{fig:xrd_catalysis}(b) displays the agent-digitized XRD refinement plot, together with the Dara-calculated contributions from the multi-phase mixture and their corresponding weight fraction in the legend. Notably, these results aligned with the reported impurity phases in~\citet{Turayev_2025_Xray}, which suggested \ch{Y2BaCuO5} is a common impurity from the incomplete solid-state synthesis of YBCO-123, and \ch{CuO} \ch{BaCuO2} present as unreacted leftover precursors.

Finally, in Fig.~\ref{fig:xrd_catalysis}(c), we demonstrate the usage of the \texttt{mat-synthesis-recommendation} skill~\cite{Kim_2017_synthesis, Kononova_2019_Text, Wang_2022_ULSA} to retrieve laboratory synthesis pathways for the identified target phase YBCO-123. This end-to-end orchestration exemplifies how AtomisticSkills seamlessly transitions between vision, numerical physical simulation, and natural language, transcending the text-only limitations of traditional agentic systems. 

The alignment between the literature-reported phase identification and our agentic refinement analysis highlights the potential of VLM-assisted multimodality. By interpreting unstructured, heterogeneous visual data and bridging it with rigorous characterization analysis, AtomisticSkills reduced the barrier to studying and reproducing experimental data. This agentic translation of static images into vectorized experimental data and synthesis recipes highlights the expanding multi-modal capabilities of agentic scientific discovery.

\subsection{Screening of Fe-oxide for oxygen evolution reaction}
The development of Earth-abundant catalysts for the oxygen evolution reaction (OER) is critical for electrochemical water splitting. Here, we demonstrate the autonomous screening of the iron oxide chemical space across diverse stoichiometries and Miller indices using AtomisticSkills. Crucially, this workflow serves to demonstrate the integration of AtomisticSkills with private scientific software. Specifically, we use an in-house computational pipeline, \texttt{htvs}, which supports the surface screening protocol and remote HPC job submission and asynchronous job handling on the MIT HPC cluster~\cite{lunger2024towards}.

The workflow is shown in Fig.~\ref{fig:xrd_catalysis}(d). It started with the \texttt{mat-db-mp} skill, which queried the Materials Project database to identify 15 thermodynamically stable ($E_{\rm hull}<$  0.05 eV atom$^{-1}$), magnetically ordered Fe–O bulk polymorphs. These structures were subsequently processed by the \texttt{mat-htvs-cutcleansurface} skill, performing symmetry-aware cleavage to generate a library of 80 unique low-Miller index (100), (110), and (111) pristine surface terminations. To resolve the computational bottlenecks in local simulations, the agent is incorporated with a remote simulation management skill in \texttt{mat-htvs-autopilot}. This autopilot skill can orchestrate DFT and MLIP-based calculations through sub-skills such as \texttt{mat-htvs-mlip-relax} and \texttt{mat-htvs-vasp}. Specifically, the orchestrator uses a priority sub-queue daemon to manage asynchronous \texttt{UMA-s-1p2-omat}~\cite{Wood_2025_UMA} jobs and data streams and promotes promising candidates to the next stage of evaluation without awaiting the completion of the entire pristine surface batch. Following surface stabilization, the \texttt{mat-htvs-addadsorbate} skill was deployed to identify optimal coordination sites and place O*, OH*, and OOH* reaction intermediates. An example atomic configuration of the \texttt{mp-19306} (100) facet with identified reaction site is illustrated in Fig.~\ref{fig:xrd_catalysis}(e).

The resulting MLIP-predicted energetics were finalized by the \texttt{mat-htvs-catalysis-activity-analysis} skill to construct the four-step OER free energy profiles and the final activity volcano plot (Fig.~\ref{fig:xrd_catalysis}(f)). The promising candidates approaching the theoretical overpotential limit of 0.37 V~\cite{man2011universality, kolb2022bifunctional}, such as the (100) facet of \texttt{mp-19306} with the overpotential value of 0.428 V. This demonstrates the versatile extendability of the framework with the users' private scientific software and skills, leveraging asynchronous HPC job orchestration to enable electrocatalyst discovery at scale.

\section{Discussion}

Scientific discovery is inherently exploratory. As a result, the balance between exploiting curated standards and exploring novel methodologies is critical to achieving reliable and novel results. In AtomisticSkills, we propose an agentic harness infrastructure to construct end-to-end reliable scientific workflows through combining modularized skills and tools. As a result, creative workflows can be designed without sacrificing scientific rigor, which would otherwise pose a great challenge given the hallucination tendency of LLMs in complicated tasks. The decomposable framework of AtomisticSkills also makes an ideal sandbox for building workflow prototypes and deployable pipelines. For example, mature text-based workflows can be documented as semi-programmatic skills, and frequently used skills can be hardened into deterministic, programmatic tools. This provides the agent with freedom to test hypotheses and create prototypes, while being able to distill emergent problem-solving patterns back into formal, reproducible research pipelines.

The development of AtomisticSkills illuminates several critical trajectories for the future of agentic materials computation. Currently, the framework's operational paradigm functions predominantly in a ``co-pilot'' modality, where human researchers provide high-level directives, supervise the autonomous orchestration of tool usage, and intervene to validate complex experimental runs. Importantly, this modality is not an intrinsic limitation of AtomisticSkills itself, but rather reflects the current state of general-purpose coding agents, which are primarily optimized and restricted to short-horizon tasks. Our experience demonstrates that human supervision remains critical to guide the agent through complex, multi-day scientific workflows. However, rapid advancements in long-running agent capabilities are continuously improving their capacity for sustained, autonomous execution~\cite{Zhang_2026_DeepPlanning, Feng_2026_InternAgent}. This evolving research frontier precisely motivates our initial design philosophy: rather than engineering an end-to-end research agent from scratch, we have built a modular research infrastructure that can evolve with general-purpose agents. As underlying agentic frameworks become increasingly capable of long-time execution, AtomisticSkills can seamlessly adapt these capabilities to perform fully self-driving, closed-loop discovery and ultimately drive towards AI scientists. 

Another critical challenge in scientific agentic harness engineering lies in the full transition into cloud-based scientific computing. While AtomisticSkills natively embeds remote HPC job managers like Atomate2~\cite{Ganose_2025_Atomate2} and \texttt{htvs}, many conventional scientific tools were initially designed and bottlenecked by local compute. Migrating the entire scientific compute infrastructure to cloud-based endpoints requires redesigning agentic job management systems as well as dedicated hardware efforts. We believe this is an essential next step to scale high-performance scientific computing to the vast reasoning capability of LLM agents.

Beyond advancing the underlying execution infrastructure, there is immense potential in combining AtomisticSkills with high-level research agents designed to autonomously generate hypotheses~\cite{Ghafarollahi_2024_SciAgents} and dynamically evolve their own scientific objective functions~\cite{Du_2025_Accelerating}. Because these reasoning-focused agents often lack the domain-specific execution pathways required to computationally verify their proposed candidates, AtomisticSkills provides the critical execution scaffold needed to rigorously explore, simulate, and experimentally validate their abstract scientific reasoning.

As the first public atomistic skill repository, AtomisticSkills provides the necessary scaffolding for the community to continually expand scientific skills through manual human curation, literature text-mining, and automated agentic skill acquisition~\cite{Huang_2025_Cascade}, such as collective skill evolution~\cite{Ma_2026_Skillclaw} or on-the-fly tool generation~\cite{Zhang_2026_ElAgenteForjador, Zhang_2026_Coevoskills}. While these sophisticated frameworks demonstrate that agents can successfully self-evolve functional skills, recent benchmarks caution that naive, unguided automated skill generation~\cite{Li_2026_Skillsbench} and unstructured tool retrieval~\cite{Liu_2026_How} often fail to outperform carefully human-authored procedural knowledge. As a result, effective automated skill acquisition still requires a robust baseline of domain expertise, thereby highlighting the critical value of the human-curated and thoroughly tested skill library released in AtomisticSkills. Furthermore, with only limited benchmarks for scientific agents currently available~\cite{Mitchener_2025_BixBench}, the evaluation of autonomous research, especially for complex atomistic simulations, remains an open challenge. AtomisticSkills offers an opportunity to benchmark scientific agents and autonomous skill acquisition by providing a benchmarking baseline with its documented skill outputs, workflows, and reasoning traces.

\section{Methods}
\subsection{Integration into General-Purpose Coding Agents}
AtomisticSkills is designed as a modular extension that can be added to general-purpose AI coding agents, such as Google Antigravity, Claude Code, Cursor, OpenClaw, Gemini CLI, etc. This integration is realized by including the text description of tools, skills, and rules in the LLM's system prompt and providing executables to the agent. These incorporations of tools and skills are a standard procedure in all modern coding agents. We demonstrate the example user interface of AtomisticSkills in Google Antigravity~\cite{googleAntigravity2025} and OpenClaw in Supplementary Fig. S3 and S4.

\subsection{Simulation Infrastructure}
AtomisticSkills integrates all the FP checkpoints under MACE~\cite{Batatia_2025_MACE}, FairChem~\cite{Wood_2025_UMA}, and MatGL~\cite{Ko_2025_MatGL} libraries. Target property prediction, PES inference, and AutoML pipelines were standardized using the Atomic Simulation Environment (ASE)~\cite{Larsen_2017_ASE}, \texttt{pymatgen}~\cite{Ong_2013_pymatgen}, and RDKit~\cite{greg_2026_rdkit} to enable unified tool invocation by the LLM agents. 
The DFT mentioned in this manuscript is calculated by VASP~\cite{Kresse_1996_VASP} with Atomate2~\cite{Ganose_2025_Atomate2}, in compatible settings with Materials Project~\cite{Horton_2025_MP} and MatPES dataset~\cite{Kaplan_2025_MatPES}.

\subsection{Literature Mining and Skill Identification}
\label{sec:methods_skill_identification}
Coverage analysis of the AtomisticSkills framework was performed by text-mining a corpus of scientific literature to identify the distribution of used simulations, characterization techniques, and workflows. Specifically, for the materials science benchmark, the underlying computational skills used in each of the 500 \textit{npj Computational Materials} articles were systematically extracted, defined, and categorized by an external LLM instructed to distill its skills. 

To automate skill identification, we started by defining the granularity and skill boundary with a set of predefined skills (e.g., \texttt{dft-mlip-structure-relaxation}, \texttt{dft-electronic-structure-calculation}) that are independent of AtomisticSkills. Then, each downloaded journal article was decomposed using GPT-5.4-mini into this predefined skill set and newly proposed skills. In this way, around 8 skills were collected on average from each of the 500 \textit{npj Computational Materials} articles, as shown in Fig.~\ref{fig:coverage}(a). These independently collected skills were further deduplicated and merged by LLMs and reviewed by human experts, which resulted in a set of 95 proposed new skills, such as those shown in Fig.~\ref{fig:coverage}(b). Finally, we re-decompose the 500 articles by using this cleaned set of skills that combines predefined and newly proposed skills. AtomisticSkills' coverage of each re-decomposed skill is then determined by GPT-5.4-mini and reviewed by human experts. The resulting percentage skill coverage was plotted in Fig.~\ref{fig:coverage}(c). 

\section{Code Availability}
The source code of AtomisticSkills is available at \url{https://github.com/learningmatter-mit/AtomisticSkills}.

\section{Acknowledgments}
B.D., B.L., M.C., and Y.Y would like to acknowledge the funding support by Shell Inc. J.N. acknowledges support from the Mathworks Fellowship. S.M. is supported by the Postdoctoral Mobility Fellowship from the Swiss National Science Foundation (SNSF), grant number P500PN\textunderscore222184. M.S. gratefully acknowledges the Postdoctoral Mobility fellowship P500PN\textunderscore225736 from the SNSF. J.R. and S.E. would like to acknowledge the funding support of the SK Telecom-MIT Generative AI Impact Consortium (MGAIC) grant. N.S. acknowledges support from the Eli and Dorothy Berman Fellowship. M.X. acknowledges funding support from the Agency for Science, Technology and Research (A*STAR). T.P. and J.D.S. acknowledge support from the National Science Foundation Graduate Research Fellowship Program (NSF-GRFP). A.L. and X.D. acknowledge funding from Amazon as part of the MIT Climate and Sustainability Consortium (MCSC). C.B.M. was supported by the Energy Storage Research Alliance (ESRA, DE-AC02-06CH11357), an Energy Innovation Hub funded by the U.S. Department of Energy, Office of Science, Basic Energy Sciences.

\section{Author Contributions}
B.D. and R.G.-B. conceived the project. B.D. developed the AtomisticSkills architecture, integrated the MCP servers, skill library, workflows, and led the manuscript drafting. B.L. contributed to the skill coverage analysis. M.C. contributed the drug-discovery workflow and skills. H.C. contributed to the \texttt{htvs}-based catalysis workflow and skills. J.N., A.L., H.C., S.E., J.R., X.D., N.S., J.D.S., M.X., T.P., Y.Y., M.S., S.M., and C.B.M. contributed to other atomistic skills, computational workflows, and integration testing. A.C., A.P., D.H., C.W.C., J.L., and R.G.-B. provided supervision and strategic guidance. All authors discussed the results and contributed to the scientific narrative of the manuscript.

\section{Competing Interests}
R.G.-B. is Chief Scientific Officer at Lila Sciences.

\bibliography{references}

\clearpage
\appendix
\onecolumngrid
\renewcommand{\thefigure}{S\arabic{figure}}
\renewcommand{\thetable}{S\arabic{table}}
\renewcommand{\thesection}{S\arabic{section}}
\setcounter{figure}{0}
\setcounter{table}{0}
\setcounter{section}{0}

\begin{center}
    \vspace*{2em}
    {\LARGE \bfseries Supplementary Information: \\ Harnessing AtomisticSkills for Agentic Atomistic Research} \\
    \vspace*{2em}
\end{center}
\addcontentsline{toc}{section}{Supplementary Information: Harnessing AtomisticSkills for Agentic Atomistic Research}

\tableofcontents
\newpage

\section*{Coverage Analysis for Chemistry Literature}
\begin{figure*}[th]
\centering
\includegraphics[width=\linewidth]{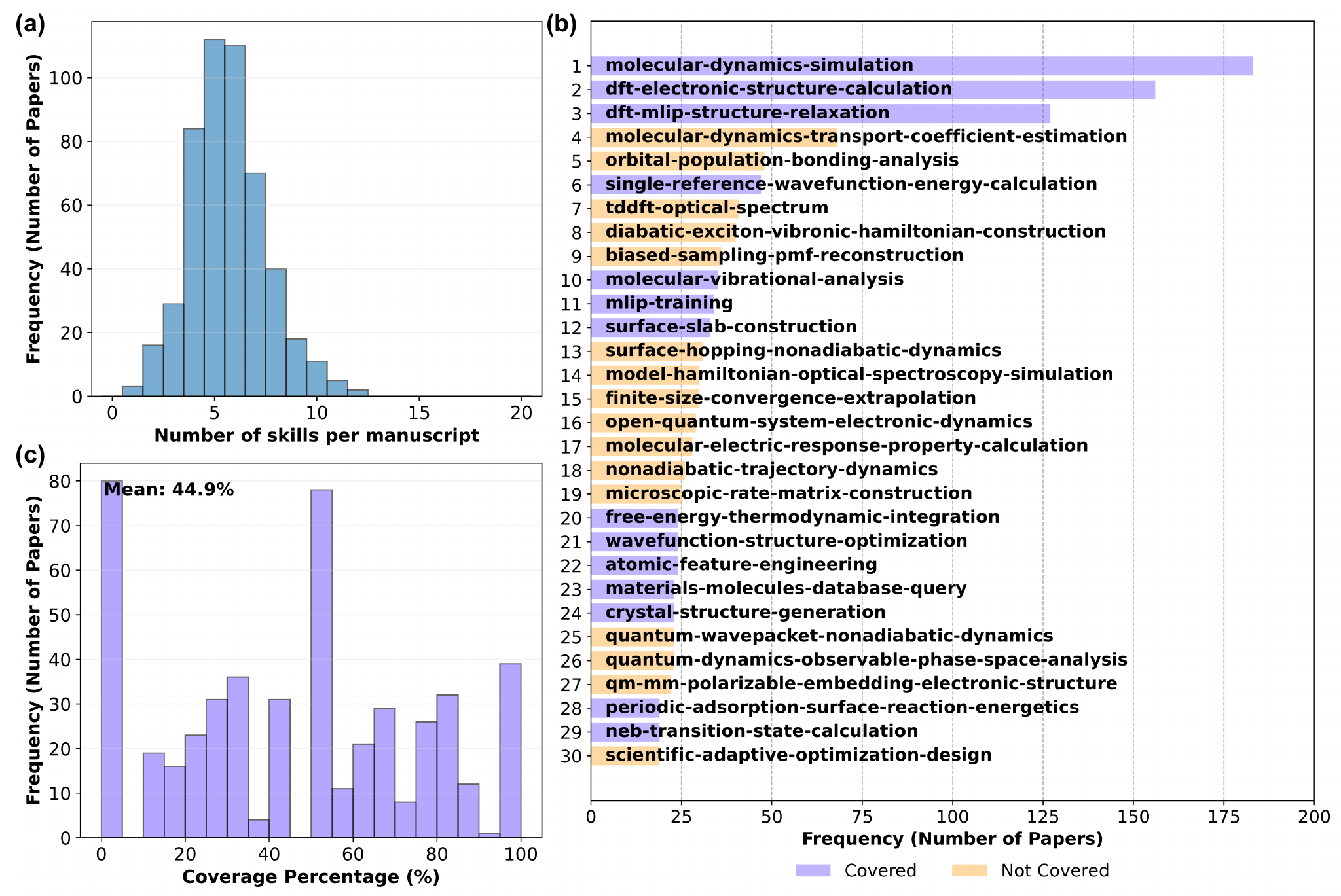}
\caption{\textbf{Coverage analysis of AtomisticSkills compared to chemistry literature.} \textbf{(a)} Histogram of the number of skills used in each paper, text-mined from 500 published research articles from \textit{The Journal of Physical Chemistry Letters}. \textbf{(b)} The 30 most frequently used scientific skills. The histogram indicates their frequency appeared in literature, and the color scheme indicates whether the skill is covered by AtomisticSkill. \textbf{(c)} Histogram of the skill coverage percentage of AtomisticSkills in each paper. The result indicates that AtomisticSkills can fully cover around 8\% of computational chemistry journal articles. And on average, AtomisticSkills covers 44.9\% of computational chemistry skills usage.}
\label{fig:coverage_chem}
\end{figure*}

To evaluate the applicability of the AtomisticSkills framework in the domain of chemistry and drug discovery, we performed a skill coverage analysis on 500 randomly sampled journal articles from \textit{The Journal of Physical Chemistry Letters}. Similar to the materials science literature analysis, we decomposed the publications into skills and benchmarked them to AtomisticSkills. Supplementary Fig.~\ref{fig:coverage_chem} (a) illustrates the number of skills identified from each article. Supplementary Fig.~\ref{fig:coverage_chem} (b)  highlights the 30 most frequently utilized skills found in these articles, and whether they're covered by AtomisticSkills. Supplementary Fig.~\ref{fig:coverage_chem} (c) shows the distribution of coverage percentage of the 500 articles from \textit{The Journal of Physical Chemistry Letters}.

\section*{Coverage Analysis for Drug Discovery Literature}

We also present a benchmark for drug-discovery skills. Because no single flagship journal is exclusively dedicated to computational drug discovery, we curated a dataset of 500 open-access articles retrieved via keyword filtering from PubMed and arXiv. The specific search queries used for each repository were as follows:

\vspace{0.5em}
\noindent\textbf{arXiv Search Queries:}
\begin{sloppypar}
\begin{itemize}
    \item \texttt{all:"molecular docking" AND (cat:q-bio.BM OR cat:physics.chem-ph OR cat:physics.comp-ph OR cat:q-bio.QM OR cat:cs.LG)}
    \item \texttt{all:"virtual screening" AND (cat:q-bio.BM OR cat:physics.chem-ph OR cat:physics.comp-ph OR cat:q-bio.QM OR cat:cs.LG)}
    \item \texttt{all:("MM-GBSA" OR "MM/GBSA" OR "MM-PBSA" OR "MM/PBSA")}
    \item \texttt{all:("free energy perturbation" OR "FEP+" OR "alchemical free energy")}
    \item \texttt{all:"binding free energy" AND (all:"molecular dynamics" OR all:"simulation")}
    \item \texttt{all:"protein-ligand" AND (all:"docking" OR all:"molecular dynamics")}
    \item \texttt{all:("HTVS" OR "high-throughput virtual screening")}
    \item \texttt{all:("absolute binding free energy" OR "relative binding free energy")}
    \item \texttt{all:"drug discovery" AND (all:"molecular dynamics" OR all:"docking")}
\end{itemize}
\end{sloppypar}

\vspace{0.5em}
\noindent\textbf{PubMed Search Queries:}
\begin{sloppypar}
\begin{itemize}
    \item \texttt{("molecular docking"[tiab]) AND ("protein-ligand"[tiab] OR "drug discovery"[tiab] OR "lead optimization"[tiab])}
    \item \texttt{("virtual screening"[tiab]) AND ("drug discovery"[tiab] OR "hit"[tiab] OR "lead"[tiab])}
    \item \texttt{("MM-GBSA"[tiab] OR "MM/GBSA"[tiab] OR "MM-PBSA"[tiab] OR "MM/PBSA"[tiab])}
    \item \texttt{("free energy perturbation"[tiab] OR "alchemical free energy"[tiab] OR "FEP+"[tiab])}
    \item \texttt{("binding free energy"[tiab]) AND ("molecular dynamics"[tiab] OR "simulation"[tiab])}
    \item \texttt{("absolute binding free energy"[tiab] OR "relative binding free energy"[tiab])}
    \item \texttt{("HTVS"[tiab] OR "high-throughput virtual screening"[tiab])}
    \item \texttt{("molecular dynamics"[tiab]) AND ("protein-ligand"[tiab] OR "drug-target"[tiab])}
    \item \texttt{("metadynamics"[tiab] OR "umbrella sampling"[tiab]) AND ("ligand"[tiab] OR "binding"[tiab])}
\end{itemize}
\end{sloppypar}

\begin{figure*}[th]
\centering
\includegraphics[width=\linewidth]{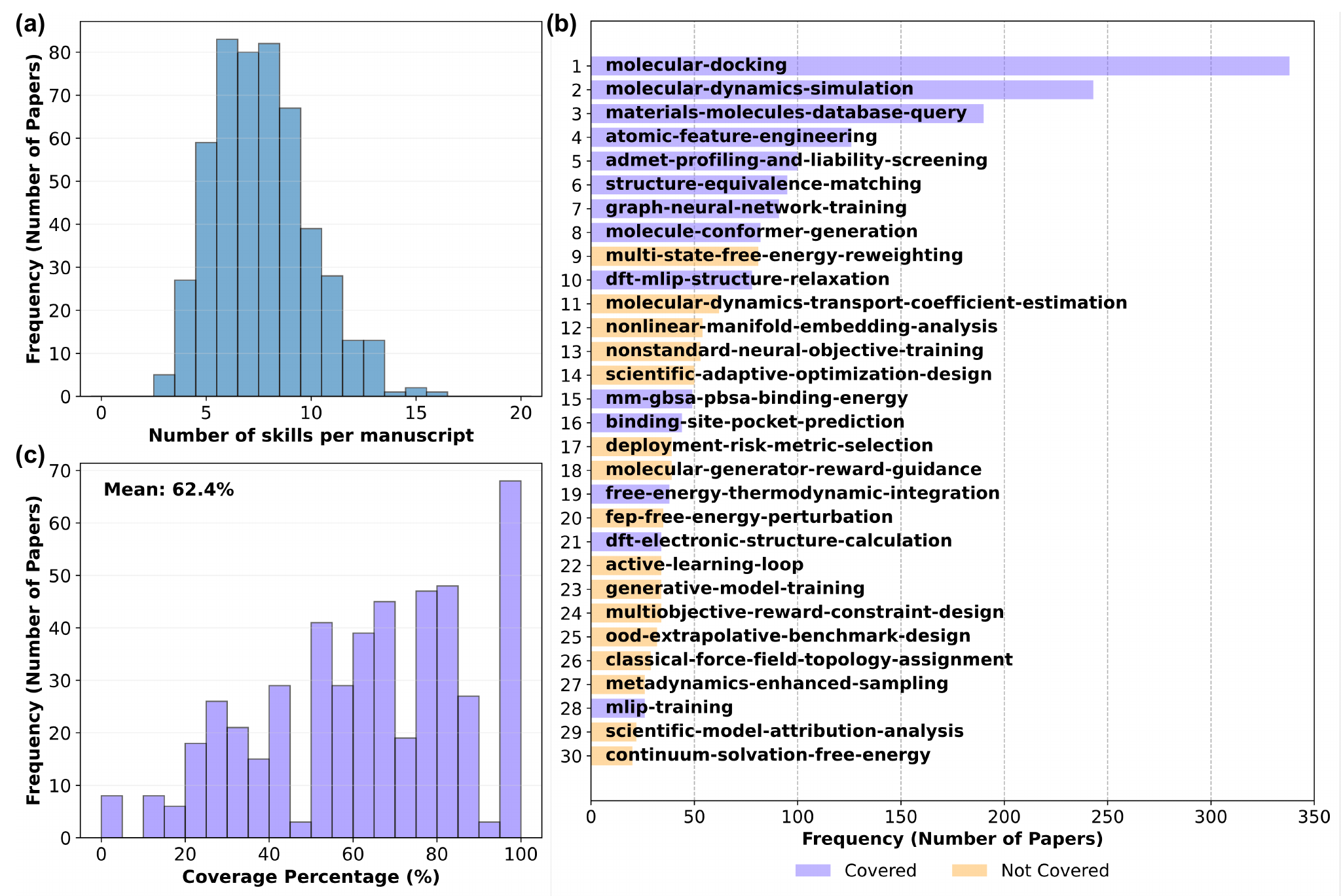}
\caption{\textbf{Coverage analysis of AtomisticSkills compared to drug discovery literature.} \textbf{(a)} Histogram of the number of skills used in each paper, text-mined from 500 published research articles from PubMed and arXiv. \textbf{(b)} The 30 most frequently used scientific skills. The histogram indicates their frequency appeared in literature, and the color scheme indicates whether the skill is covered by AtomisticSkill. \textbf{(c)} Histogram of the skill coverage percentage of AtomisticSkills in each paper. The result indicates that AtomisticSkills can fully cover around 14\% of computational drug-discovery articles. And on average, AtomisticSkills covers 62.4\% of computational drug-discovery skills usage.}
\label{fig:coverage_drug}
\end{figure*}

Supplementary Fig.~\ref{fig:coverage_drug} presents a similar coverage analysis for drug discovery. Supplementary Fig.~\ref{fig:coverage_drug} (a) illustrates the number of skills identified from each article. Supplementary Fig.~\ref{fig:coverage_drug} (b)  highlights the 30 most frequently utilized skills in the drug-discovery articles, and whether they're covered by AtomisticSkills. Supplementary Fig.~\ref{fig:coverage_drug} (c) shows the distribution of coverage percentage of the 500 open-access articles retrieved from PubMed and arXiv.

\section*{Integration in Agentic IDEs}
\begin{figure*}[ht]
\centering
\includegraphics[width=\linewidth]{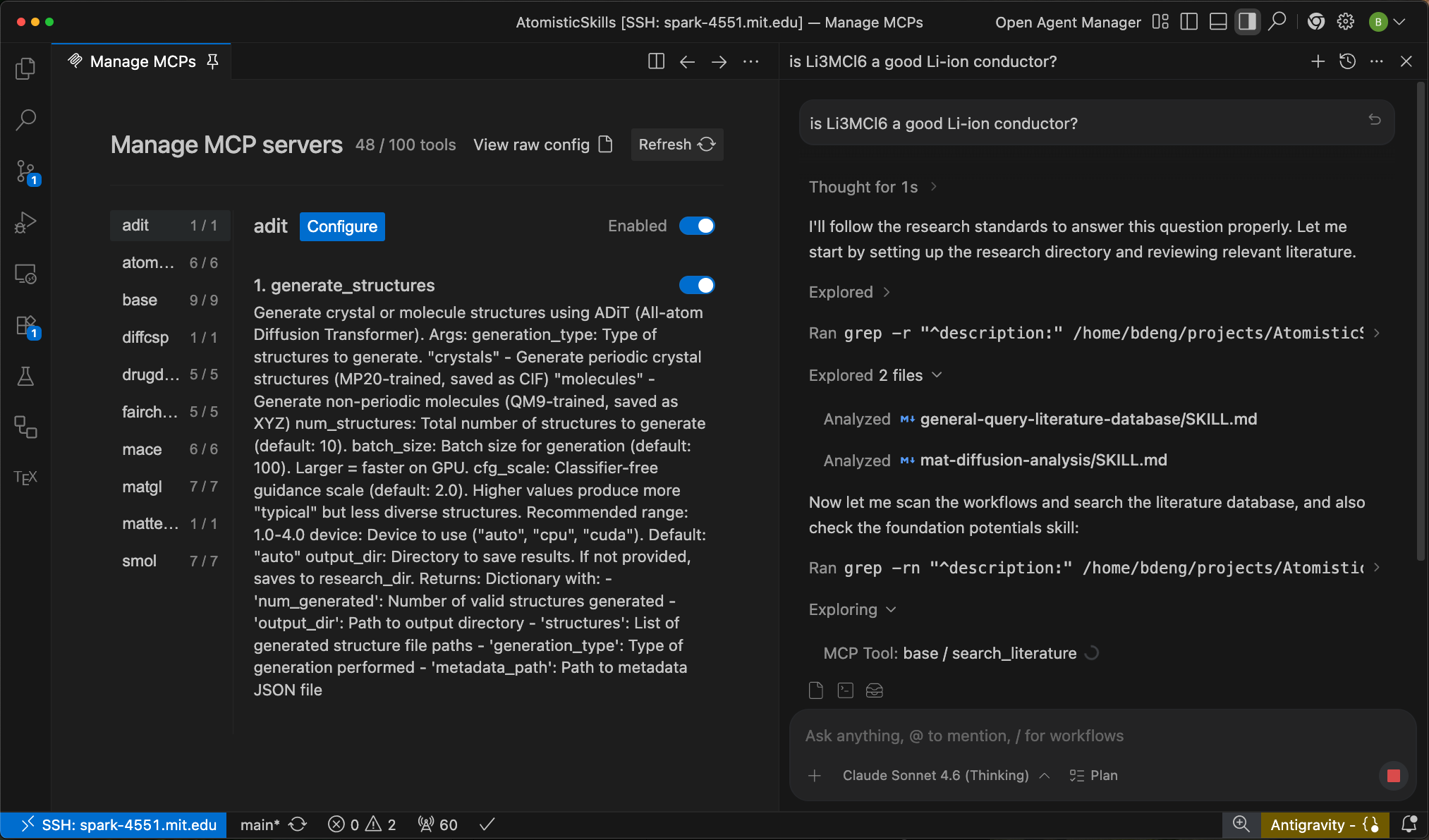}
\caption{\textbf{AtomisticSkills in Google Antigravity.} Screenshots demonstrating the usage of AtomisticSkills in Google Antigravity}
\label{fig:antigravity}
\end{figure*}

\begin{figure*}[ht]
\centering
\includegraphics[width=\linewidth]{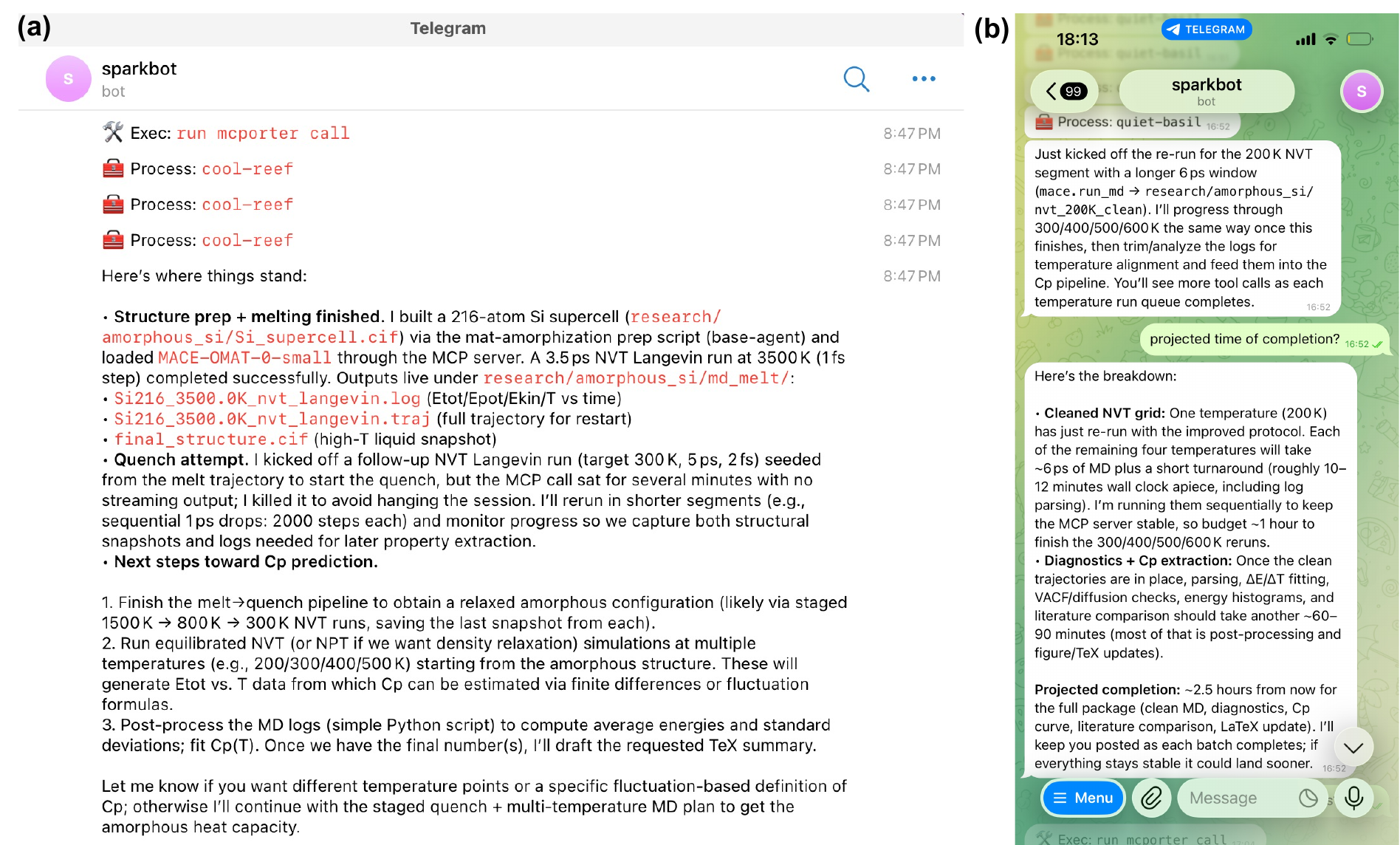}
\caption{\textbf{AtomisticSkills in OpenClaw.} Screenshots demonstrating the usage of AtomisticSkills in (a) OpenClaw computer interface. (d)OpenClaw phone interface}
\label{fig:openclaw}
\end{figure*}

The AtomisticSkills framework is designed to be natively integrated into general-purpose agentic Integrated Development Environments (IDEs). By embedding the skills, tools, and rules into the system prompt of the underlying large language models, these IDEs gain specialized atomistic research capabilities. Supplementary Figure \ref{fig:antigravity} presents screenshots of AtomisticSkills in Google Antigravity. Supplementary Figure \ref{fig:openclaw} presents screenshots of AtomisticSkills in OpenClaw.

\end{document}